\documentclass[11pt,preprint,flushrt]{aastex631}

\usepackage{natbib,graphicx,amsmath,subfigure,color}
\usepackage{verbatim}
\usepackage{soul} 

\def\eqq#1{Equation~(\ref{#1})}

\newcommand\eg{{\it e.g.\/}}
\newcommand\ie{{\it i.e.\/}}

\newcommand\px{Planet X}
\newcommand\lsst{\textit{LSST}}
\newcommand\gaia{\textit{Gaia}}

\newcommand{\uas}{\mbox{$\mu\textrm{as}$}}
\newcommand{\rhohat}{\mbox{\boldmath $\hat \rho$}}
\newcommand{\phihat}{\mbox{\boldmath $\hat \phi$}}
\newcommand{\zhat}{\mbox{\boldmath $\hat z$}}
\newcommand{\shat}{\mbox{\boldmath $\hat s$}}
\newcommand{\nhat}{\mbox{\boldmath $\hat n$}}
\newcommand{\ohat}{\mbox{\boldmath $\hat o$}}
\newcommand{\yhat}{\mbox{\boldmath $\hat y$}}

\newcommand\change[1]{{#1}}
\begin{document}


\keywords{Planetary dynamics, Fisher's Information, Kuiper Belt, Trans-neptunian objects, Astrometry, Gravitation}
\title{Can the gravitational effect of Planet X be detected in current-era tracking of the known \change{major and minor} planets?}

\author[0000-0001-6299-2445]{Daniel C. H. Gomes}
\email{dchgomes@gmail.com}
\affil{Department of Physics \& Astronomy, University of Pennsylvania, 
209 S.\ 33rd St., Philadelphia, PA 19104}

\author[0000-0002-8076-3854]{Zachary Murray}
\affil{Center for Astrophysics | Harvard \& Smithsonian, 60 Garden Street, Cambridge, MA 02138}

\author[0000-0002-3800-5662]{Rafael C. H. Gomes}
\affil{Department of Physics \& Astronomy, University of Pennsylvania, 
209 S.\ 33rd St., Philadelphia, PA 19104}

\author[0000-0002-1139-4880]{Matthew J. Holman}
\affil{Center for Astrophysics | Harvard \& Smithsonian, 60 Garden Street, Cambridge, MA 02138}

\author[0000-0002-8613-8259]{Gary M. Bernstein}
\affil{Department of Physics \& Astronomy, University of Pennsylvania, 
209 S.\ 33rd St., Philadelphia, PA 19104}

  \begin{abstract}
    Using Fisher information matrices, we forecast the uncertainties $\sigma_M$ on the measurement of a ``Planet X''  at heliocentric distance $d_X$ via its tidal gravitational field's action on the known planets.  Using planetary measurements currently in hand, including ranging from the \textit{Juno, Cassini,} and Mars-orbiting spacecraft, we forecast a median uncertainty (over all sky positions) of $\sigma_M=0.22M_\oplus (d_x/400\,\textrm{AU})^3.$  A $5\sigma$ detection of a $5M_\oplus$ Planet X at $d_X=400$~AU should be possible over the full sky but over only 5\% of the sky at $d_X=800$~AU. The gravity of an undiscovered Earth- or Mars-mass object should be detectable over 90\% of the sky to a distance of 260 or 120~AU, respectively. Upcoming Mars ranging improves these limits only slightly.  We also investigate the power of high-precision astrometry of $\approx8000$ Jovian Trojans over the 2023--2035 period from the upcoming \textit{Legacy Survey of Space and Time (LSST).}  We find that the dominant systematic errors in optical Trojan astrometry (photocenter motion, non-gravitational forces, and differential chromatic refraction) can be solved internally with minimal loss of information.  The Trojan data allow cross-checks with Juno/Cassini/Mars ranging, but do not significantly improve the best-achievable $\sigma_M$ values until they are $\gtrsim10\times$ more accurate than expected from \lsst.
    The ultimate limiting factor in searches for a \px\ tidal field is confusion with the tidal field created by the fluctuating quadrupole moment of the Kuiper Belt as its members orbit.
This background will not, however, become the dominant source of uncertainty until the data get substantially better than they are today.
\end{abstract}

\section{Introduction}

State-of-the-art \change{ranging and astrometry} of solar system bodies has yielded several of the
greatest discoveries in the history of astronomy and physics, including the development of laws of planetary motion \citep{Kepler12,Kepler3} and general motion \citep{Principia}, the successful prediction of Neptune's existence from anomalies in Uranus's orbit \citep{LeVerrier}, the precise measurement of the astronomical unit~\citep{AUradar,Smith_1966,Ash_1966}, and the validation of General Relativity \citep{Einstein,Shapiro_1968}.  Spectacular mis-interpretations of the data include the prediction of Planet Vulcan \citep{Vulcan}. We use a Fisher matrix formulation to estimate the sensitivity of historical \change{positional/astrometric} measurements of the major planets, plus data expected in the next decade,  to discover the gravitational signature of unknown bodies in the solar system, \change{which we will generically refer to as ``\px,'' an appellation first used by \citet{Dallet1901} for a hypothesized trans-Neptunian planet. There has in particular been attention in recent years to a potential Neptune-mass \px\  lurking hundreds of AU from the Sun, sometimes referred to as ``Planet 9.''  But there is also some theoretical motivation for, and weak observational limits on, Earth- or Mars-mass bodies at closer distances \citep[\eg][]{Volk2017, Silsbee2018}, and the gravitational constraint forecasts in this work are equally applicable to these as \px. }

\subsection{Searching for \px}

Although the prediction and discovery of Neptune motivated searches for additional planets~(see~\citealt{Tremaine_1990} and \citealt{Batygin_2019} for reviews), our census of the solar system remained relatively static for many decades.  Tombaugh's 1930 discovery of Pluto~\citep{Tombaugh_1946,Tombaugh_1996} marked the beginning of a continuing, almost fantastical era of outer solar \change{system} discovery, including Oort's hypothesis of a nearly spherical \change{shell} of comets orbiting our sun at great distances~\citep{Oort_1950}; the discovery of the first Centaurs~\citep{Kowal_1979} and the realization that there must be additional reservoirs supplying such a dynamically short-lived population~\citep{Duncan_1988,Holman_1993}; the discovery of 1992~QB1, the first trans-neptunian object~(TNO)~by Jewitt and Luu~\citep{Jewitt_1993}; and the recognition of the intricate dynamical structure of the small body population beyond Neptune and what it implies for the formation of the solar system~(see \citealt{Gladman_2021} for a review). Of course, this era includes the unbelievably successful exploration of our solar system by spacecraft encountering, orbiting, landing on, and crashing into members of our solar system, large and small.

All the while, the hypothesis of additional massive planets, either at an earlier era or still residing in the outer solar system, continues to be compelling.  Supporting evidence primarily comprises features in the orbital structure in the outer solar system that can be explained by the long-term dynamical influence of an additional planet~\citep{Gladman_2002, Brunini_2002, Melita_2004, Gladman_2006, Lykawka_2008,Huang_2022}. Most recently, the orbital distribution of trans-neptunian objects (TNOs) with highly eccentric orbits and large perihelion distances has motivated new searches~\citep{Trujillo_2014,Sheppard_2016,Batygin_2016,Brown_2016}.
Attempts to discover \px\, have followed two primary approaches: searching for the light reflected or emitted by the planet itself; and searching for the signatures of the short-term gravitational influence of \px\, on other bodies in the solar system.

Discovering \px\, directly requires large portions of the sky to be carefully surveyed to faint magnitudes,
unless dynamical or other evidence can narrow the search region~\citep{Brown_2016,Malhotra_2016}.  Only a few telescopes have the combination of aperture and field of view required to support an efficient search over a large area of sky.  Furthermore, the search algorithms and survey cadence used must be sensitive to the very slow rates of apparent motion for distant solar system bodies observed from Earth.  Nearly every platform with which a meaningful search could be conducted has been used or considered, including Spacewatch~\citep{Larsen_2007}, the Catalina Sky Survey~\citep{Brown_2015}, Palomar~\citep{Brown_2022}, Pan-STARRS~\citep{Holman.2018}, the Canada-France-Hawaii telescope~\citep{Bannister.2018}, DECam~\citep{Bernardinelli_2022a,Bernardinelli_2022b}, NASA's IRAS~\citep{Beichman_1988}, WISE~\citep{Luman_2014} and TESS missions~\citep{Holman_2019,Rice_2020}, and even CMB experiments~\citep{Baxter_2018,Naess_2021}.

Many claims of the existence of additional unseen planets, based on trends in the residuals of the observations against the available ephemerides, have been made over the years.  However, \citet{Hogg_1991} demonstrated in a series of numerical experiments that such trends are to be expected when an orbital solution is fit to observations spanning less than a full period and then extrapolated to the times of new observations without incorporating the new data.  The trends vanish when the orbit is re-fit using the full set of observations.  Indeed, \citet{Standish_1993} dispatched all claims existing at the time by showing that the trends seen at the time disappear when improved masses of the giant planets (based on Voyager data) are used and the orbits re-fit.  \citet{Hogg_1991} presciently concluded in their investigation that continued searches for \px\, would not be well motivated until improved data on the outer planets, from Cassini or Galileo or pulsar timing, became available.  

In an inspiring investigation, \citet{Fienga_2016} used just such data, a long span of Cassini ranging observations, to constrain the mass and location of \px.  The orbit for \px\ in \citet{Batygin_2016} included all the elements except for the true anomaly, or its equivalent, because the dynamical influence on the extreme TNOs was argued to be primarily secular.  By inserting a synthetic \px\ with an assumed mass of $10~M_\oplus$, carrying out full ephemeris solutions to all the available data, and comparing the residuals to those without the addition of \px, \citet{Fienga_2016} were able to show that a broad range of initial longitudes is excluded because the residuals would be excessive.  Furthermore, the residuals would be significantly improved by the addition of a large planet at other initial longitudes.  \citet{Holman_2016a} extended this work to include a wide range of masses and initial positions of the planet across the entire sky.  The \citet{Fienga_2016} investigation is based on an earlier version of the INPOP ephemeris model that did not include mass in the Kuiper belt aside from Pluto~\citep{Fienga_2016b}, and the \citet{Holman_2016a} study is built upon the residuals shown in \citet{Fienga_2016} and therefore inherits most of its dynamical properties.  As demonstrated by \citet{Pitjeva_2018}, however, incorporating the mass of the Kuiper belt, through a combination of objects with known masses and orbits and massive rings to account the gravitational influence of a distribution of smaller bodies, reduces the residuals in the ephemeris fits.  In particular, it reduces the residuals in the Cassini ranging observations.  \citet{fienga2020} reprised the investigation using the INPOP19a model, which includes the gravitational influence of the Kuiper belt, and a tidal model as in \citet{Holman_2016a}.  Their results exclude an additional planet \change{within $\sim25\arcdeg$ from the ecliptic plane,} closer than $500~AU$ if it is $5~M_\oplus$ or closer than $650~AU$ if it is $10~M_\oplus.$

Following the approach of~\citet{Hogg_1991}, \citet{Holman_2016b} tested
whether fits to the orbits of Pluto and a set of well-observed TNOs would be improved by the inclusion of a \px.  They focused on the data set developed and used by \citet{Buie_2015} to refine the orbit of the Pluto-Charon in preparation for the {\it New Horizons} encounter.   The results of \citet{Holman_2016b} suggest a planet that is either more massive or closer than the best-first results of \citet{Batygin_2016} or \citet{Fienga_2016}.   This result was driven by a clear trend in the residuals over the course of two decades which may be due to systematic errors.  Alternatively, it could also be explained by perturbations from a different body, aside from \px, that is
closer to Pluto but less massive than $5-10~M_\oplus$.

In this work, we carefully examine the power of extant and imminent observations
(astrometry, ranging, and masses derived from satellites) to constrain the mass of \px, while allowing for freedom in the masses and orbits of all known significant gravitating bodies in the solar system.  We do this as a function of all
possible \px\ locations in the sky.

\subsection{A new era for ground-based astrometry}
In the past 50 years, active radio ranging to spacecraft orbiting or flying by planets and radar ranging to the inner planets have largely superseded ground-based optical astrometry for high-precision ephemeris constraints.  Radio ranging to the terrestrial planets routinely reaches sub-meter accuracy, and lunar laser ranging attains mm-level accuracy.  In the search for a gravitational signature for \px, however, the tidal acceleration $a_{\rm tidal}$ increases linearly with heliocentric distance $d_T$ of the test body, and the time over which the perturbations oscillate is the period $P,$ leading some signals to scale as $a_{\rm tidal}P^2 \propto d_T^4$ (see Section~\ref{sec:quadsize}).   Hence, there is a signal-to-noise ($S/N$) premium on observations of bodies in the outer solar system, as well as the advantage of reducing the influence of potentially confounding signals from uncertainties in the mass distribution of the asteroid belt.   But spacecraft ranging to the outer planets is much sparser, with meter-level precision to Jupiter (\textit{Juno}) and Saturn (\textit{Cassini}) and only single Voyager flybys to Uranus and Neptune.  Our first focus will be to assess the power of past and ongoing ranging observations (primarily of Mars, Jupiter, and Saturn) to constrain \px.

Our second focus is to assess the power of traditional ground-based optical astrometry to augment state-of-the-art ranging data in constraining \px.  Ground-based astrometry is currently undergoing a revolution in both \emph{quality} and \emph{quantity} of available data.
The per-exposure astrometric accuracy of routine ground-based imaging has been improved to milliarcsecond (mas) level by three developments:
\begin{itemize}
\item \citet{decamast} demonstrate $\approx2$~mas accuracy in mapping the 500-Mpix focal plane of the \textit{Dark Energy Camera} to (relative) sky coordinates.  This is done primarily by solving for internal consistency of stellar positions on multiple shifted exposures of the same star field.
\item The availability of a dense global network of stars with $\approx20$~\uas\ absolute astrometric accuracy from the \gaia\ spacecraft allows one to tie any ground-based exposure to the absolute reference frame with very high accuracy.
\item For bright point sources \change{($m_r\lesssim20,$ depending on the observatory),} the dominant astrometric error from the ground is then stochastic refraction from the turbulent atmosphere.  \citet{fortino} demonstrate that turbulence components on scales $\gtrsim1\arcmin$ can be mapped and removed from each exposure by fitting a Gaussian Process model to the deviations of \gaia\ stars from their published positions.  This reduces the RMS turbulence signal to $\approx 2$~mas per axis.
\end{itemize}

The revolution in data \emph{quantity} is best exemplified by the upcoming 10-year \textit{Legacy Survey of Space and Time (LSST)} to be conducted from the 8-meter telescope at the \textit{Vera Rubin Observatory}.  There are good reasons to expect the performance of the \lsst\ camera to be better than that of the \textit{Dark Energy Survey}.  We can therefore plausibly expect the floor on astrometric errors of \lsst\ observations of unresolved minor planets to be $\sigma_{\rm min} \approx 2$~mas.  This is better than the \gaia\ single-epoch uncertainty for sources with $G>18$ \citep{gaiadr3ss}.  

But \lsst\ will observe essentially \emph{every} minor planet at declination $<30\arcdeg$ nearly 1000 times in its 10-year survey, including for example \change{$\sim8000$  Jovian Trojan asteroids at $H < 14.5$. As detailed in Section~\ref{sec:trojans}, the collective accuracy on a shift of this entire population is $\approx2.5\uas$, which is 8~meters at the distance of the Trojans.} This is comparable to the accuracy obtained by the \textit{Cassini} and \textit{Juno} spacecraft.  \lsst's collective precision on the main-belt asteroid (MBA)  population will be at the sub-meter level, given that they are $2\times$ closer, and $O(10^5)$ of them will be bright enough to approach $\sigma_{\rm min}.$ 

The tracking of many minor planets can have advantages over the \textit{Cassini} and \textit{Juno} ranging. First, ranging is only sensitive to displacements in the radial direction (namely the invariable plane), whereas astrometry measures 2 transverse displacements. Second, the minor planets have larger eccentricities and a wide range of inclinations, which enhances the types of perturbations (\eg\ a polar \px) that are detectable.  Third, the Trojans and MBAs are distributed in space and we can track each for $\ge1$ full orbit, which substantially reduces degeneracies between potential sources of gravity. \textit{Cassini} data are available for only 45\%  of  Saturn's orbital period.

Note that ranging to Mars has much of these latter two advantages while also having accuracy at cm levels---so it is interesting to ask which combinations of Mars ranging, Jupiter$+$Saturn ranging, and \lsst\ Trojan tracking yield the strongest constraints.

This work is inspired by the proposal by \citet{RiceLaughlin} to measure the \px\ gravitational perturbation via a USA-spanning network of several thousand small telescopes monitoring occultations of \gaia\ stars by Jovian Trojans.  The occultation method offers higher precision and reduced systematic errors relative to traditional astrometry.  It is of great interest, however, to ask what can be achieved---at zero marginal cost in instrumentation---from the observations to be obtained by \lsst, and to investigate more rigorously the potential confusion between \px\ and other solar system gravitational anomalies.

Ground-based astrometry of minor planets has some systematic errors that do not exist for spacecraft tracking, \eg\ differential chromatic refraction (DCR) in the atmosphere, photocenter motion (PCM), and non-gravitational forces.  We will investigate how much these systematic errors degrade the potential statistical accuracy of the \lsst\ Trojan tracking.

\subsection{Plan of this paper}
We begin in Section~\ref{sec:bote} by reviewing some analytic expressions for the orbital perturbations induced by the gravity of a \px, to understand the size and scaling of the signals of interest.  In Section~\ref{sec:math}
we describe how the Fisher matrix is used to forecast the expected uncertainties $\sigma_M$ in the mass $M_X$ of a putative \px. We describe the model of our solar system used to derive these limits, including some unresolvable degeneracies between \px\ and the tidal fields generated by Kuiper Belt members.  In Section~\ref{sec:numerics} we present the numerical methods used to acquire the derivatives of observations with respect to parameters required for the Fisher method.   In Section~\ref{sec:results} we present the results of the Fisher analysis in terms of expected accuracy $\sigma_M$ on the mass of \px\ vs its location in the sky, using completed and planned observations of the major planets.  We quantify how the
addition of \lsst\ observations of Trojans affect the constraints on \px\ mass, including a formulation of the major systematic errors, in Section~\ref{sec:trojans}.  In Section \ref{sec:trojanresults} we apply our Fisher matrix results  to scenarios including \lsst\ Trojans, quantifying the accuracy on \px\  and our ability to distinguish systematic errors in the data from the signature of \px.  We conclude in Section~\ref{sec:conclusions}.  The appendices give the details of some aspects of the calculation.

\section{The back of the envelope}
\label{sec:bote}
Before solving numerically for the sensitivity $\sigma_M$ to the mass of \px, it is useful to have a rough estimate of the sizes of displacements that the presence of \px\ can cause on decade time scales. We will assume that we are tracking some test body at heliocentric distance $d_T$ with radius $R_T$, and that this is being perturbed by an effectively stationary (over 50 years) point mass $M_X$ at distance $d_X.$
\citet{Brown_2021} use the dynamics of extreme TNOs to suggest $M_X\approx 6M_\oplus$ in an orbit ranging from roughly 300--500~AU heliocentric distance. 
For nominal values we will take $d_T=5.2$~AU, $M_X=5 M_\oplus,$ and $d_X=400$~AU, \change{corresponding to the effect of a \px\ in the center of their range as influencing Jupiter or its Trojans.  The choice of nominal value is simply to scale the constraints into a useful range, not a statement on the likelihood of or constraints on a \px.}
We describe \px's position by its ecliptic azimuthal angle $\phi_X$ and polar angle $\theta_X$.  It will also be useful to define the radial, azimuthal, and vertical directions relative to the ecliptic as \rhohat, \phihat, and \zhat, respectively.  \citet{Brown_2021} derive favored regions of the sky, but we remain agnostic about \px's location and derive constraints for all sky locations, \change{so that our forecasts remain generally applicable to any \px\ hypothesis.}

\subsection{The signatures of \px}
We can decompose the gravitational field of \px\ into spherical harmonics about the solar system barycenter with radial dependences $r^\ell$.
The $\ell=0$ and $\ell=1$ terms are a potential shift and a constant acceleration, respectively, which are unobservable with differential measurements within the solar system.  The $\ell=2$ tidal terms are a sufficient description of the observable effects of \px, since $\ell\ge3$ terms are suppressed relative to this by factors of $d_T/d_X\lesssim0.01.$  We can thus rephrase our search for \px\ as a search for unexplained $\ell=2$ potential terms.  There are five real degrees of freedom in the tidal field, which we can express as the five terms in the traceless, symmetric $3\times3$ tidal matrix, or as the real (for $m\ge0$) and imaginary (for $m>0$)  coefficients $A_{2m}$ of the tidal potential
\begin{equation}
  \psi(r,\theta,\phi) = \frac{4\pi GM_Xr^2}{5 d_X^3} \sum_{m=-2}^{2} A_{2m}(\theta_X, \phi_X) Y_{2m}(\theta, \phi).
  \label{eq:tidal}
\end{equation}

A single point source like \px\ has only three degrees of freedom (\eg\ $M_X, \theta_X, \phi_X$) because the principal components of its tidal matrix must be in the ratio $-1:-1:2.$

Our method will be to construct the Fisher matrix for the 5 terms of the tidal field.  Then the perturbation for a \px\ in any posited direction $(\theta_X, \phi_X)$ can be expressed as a linear superposition of these, and we can infer a bound on $M_X/d_X^3$ for this position.  The Fisher matrix of tidal spherical harmonics is equivalent to that derived for \px s placed at five distinct positions on the sky, so we opt for computing the latter \change{because it is more convenient to obtain derivatives with respect to point masses than quadrupole coefficients within the dynamical codes.} In Appendix~\ref{sec:tidal} we show how this matrix is used to place limits on $M_X$ over the full sky.  Because all of the gravitational effects of \px\ scale as $M_X/d_X^3,$ we will quantify all our results as the measurement accuracy $\sigma_M$ attained on $M_X$ at fixed nominal distance $d_X=400$~AU.  Then we can rescale $\sigma_M$ to any distance. In particular, we will consider a measurement to be ``conclusive'' for the presence of \px\ if it can detect $M_X=5M_\oplus$ at $d_X=800$~AU at $5\sigma$ significance, which
is equivalent to obtaining $\sigma_M<0.125M_\oplus$ at the nominal distance.

The critical parameter describing the shifts, precessions, and extraneous oscillations that the tidal field generates is the ratio of the tidal acceleration to the solar acceleration:
\begin{equation}
  \epsilon \equiv  \frac{M_Xd_T^3}{M_\odot d_X^3} \approx 3\times 10^{-11} \left( \frac{d_T}{5.2\,\textrm{AU}}\right)^3
  \left( \frac{M_X}{5M_\oplus}\right)
  \left( \frac{400\,\textrm{AU}}{d_X}\right)^3.
 \label{epsilon}
\end{equation}
We are clearly justified in calculating our derivatives as first-order perturbations under this parameter.  To give a sense of scale, the total displacement over an  observing period $t$ for constant tidal acceleration for the nominal \px\ is
\begin{equation}
  \Delta x \approx \frac{1}{2} a_{\rm tidal} t^2 =  \frac{1}{2} \epsilon a_\odot t^2 \approx 500\,\textrm{m} \, \left(\frac{d_T}{5.2\,\textrm{AU}}\right)
     \left(\frac{t}{12\,\textrm{yr}}\right)^2
  \end{equation}
  and the observed angular shift if this is transverse to the line of sight from the barycenter is independent of the tracer's distance:
\begin{equation}
  \Delta \theta =  \frac{\Delta x}{d_T} \approx 130 \, \uas
     \left(\frac{t}{12\,\textrm{yr}}\right)^2.
  \end{equation}
  These maximal signals should be detectable at $S/N=$10--100 from either radio ranging observations of planets (if radial) or the \change{expected} minor-planet optical astrometry. 

The constant-acceleration model is, however, only applicable to distant bodies which traverse a small fraction of an orbit during the \lsst\ campaigns,  \eg\ trans-Neptunian objects (TNOs).  It is only roughly applicable to the \textit{Cassini} data, which span about 45\% of one Saturnian period.  But it is very difficult to disambiguate a \px\ tidal signal from other gravitational influences using such localized measurements.  Using more bodies improves our ability to constrain the spatial/temporal variation, and hence the source, of a gravitational anomaly. 
  
For data spanning one or more periods of the tracer particles, it is easier to understand the influence of the tidal fields as a combination of several effects enumerated below.  These grow over time more slowly than $t^2$ and have different scalings with target heliocentric distance $d_T.$

\subsubsection{Radial displacement}
\label{sec:displacement}
One manifestation of the $A_{20}$ field is a perturbation to Kepler's 3rd Law, \ie\ an offset in radius at fixed orbital period $P$ or mean angular motion $\Omega_0.$  For a nominal \px\ at an ecliptic pole, this is
  \begin{equation}
    \delta r = \frac{\epsilon d_T}{3} \approx 9\,\textrm{m}\, \left(\frac{d_T}{5.2\,\textrm{AU}}\right).
\label{eq:radial}
    \end{equation}
    While this distance shift is well above the uncertainties in ranging for Mars or the outer planets, it requires determining the period \change{of the monitored object} to equally high precision.  Radial shifts are observable from astrometry only through the reflex of Earth's motion, which suppresses the signal by ${(1\,\textrm{AU})/d_T},$ making this likely undetectable from \lsst\ astrometry. \change{We can alternatively view this effect as changing the period associated with a fixed semi-major axis, by an amount $\epsilon P / 2.$  This would lead to an apparent angular lag that grows linearly with time, equal to}
      \begin{equation}
        \Delta \phi = 2\pi \, \frac{\epsilon}{2} \frac{t}{P} \approx 20\uas  \left(\frac{tP}{(12\,\textrm{yr})^2}\right) 
  \left( \frac{M_X}{5M_\oplus}\right)
  \left( \frac{400\,\textrm{AU}}{d_X}\right)^3.
\label{eq:lag}
\end{equation}
\change{If the distance for a body is determined to sub-meter precision by ranging, then the period shift would eventually become detectable with \uas-level astrometry.  But we are not aware of any impending observations that would yield both ranging and astrometric data of the required precision on the same body.}

\subsubsection{Precession}
\label{sec:precession}
Another manifestation of the tidal field is a shift of the epicyclic frequencies away from the orbital frequency, generating a precession of the ascending node (for the \zhat\ epicycle, largely undetectable by ranging) and longitude of perihelion.  The latter is accessible to ranging via $\rhohat$ epicycles, and also to astrometry via $\phihat$ epicycles, which have amplitudes typically twice as large (in meters) as the radial oscillations (an advantage for astrometry over ranging).  The typical precession rate is $\epsilon \Omega_0,$ which leads to perturbations whose amplitudes grow linearly over multiple orbital cycles, leading to transverse astrometric displacements \change{(angular shifts in any direction)} of
\begin{equation}
    \Delta \theta \approx \sin(\Omega_0 t) \left\{ \begin{array}{llll}
                                                    \sin{i}\, \epsilon \Omega_0 t & \approx 43\,\uas\,\sin{i}&  \left(\frac{d_T}{5.2\,\textrm{AU}}\right)^{3/2}
                                                                                  \left(\frac{t}{12\,\textrm{yr}}\right)  & \textrm{(nodal precession)} \\
                                                    2e\, \epsilon \Omega_0 t & \approx 86\,\uas\, e & \left(\frac{d_T}{5.2\,\textrm{AU}}\right)^{3/2}
                                                                              \left(\frac{t}{12\,\textrm{yr}}\right)  & \textrm{(apsidal precession)} \\
                                                  \end{array}\right.
        \end{equation}
The signal accessible to ranging is primarily the \rhohat\ component of apsidal precession
\begin{equation}
    \Delta \rho \approx \sin(\Omega_0 t) \,e\,  \epsilon \Omega_0 t d_T \approx 160 \, \textrm{m}\, e\, \left(\frac{d_T}{5.2\,\textrm{AU}}\right)^{5/2}
    \left(\frac{t}{12\,\textrm{yr}}\right).
  \end{equation}
  These precession signals all grow in amplitude toward the outer solar system; the choice of which minor planet populations are most useful for detecting \px\ then becomes a compromise between the increasing signal and the smaller number of bright targets (higher measurement noise) as we move outwards. \change{There are $\sim1200$ Trojans with typical projected \lsst astrometric accuracy below 5~mas per component per observation, and $1.6 \times 10^5$ MBAs that would satisfy the same criterium. The $130\times$ larger number of targets among the MBAs is partly offset by the $2^{3/2}$ larger precession signal of the Jovian Trojans. We will investigate the value of the Jovian Trojans for constraints on \px, but it is possible that the MBAs will offer stronger constraints if the systematic errors associated with smaller bodies---namely radiation pressure---can be controlled. We defer investigation of the MBAs to a future publication.}
  
Jupiter and Saturn have very regular orbits ($e\approx0.05,$ $\sin i<0.02$ from the invariable plane) whereas the Trojans have an RMS eccentricity of 0.08 and an RMS $\sin i$ of 0.28, more than $10\times$ larger than Jupiter and Saturn.  This makes the precession signals from a polar \px\ much larger in this minor body population.  Mars' eccentricity of 0.093 is also more favorable to observations of apsidal precession than those of Jupiter and Saturn.

\subsubsection{Quadrupole excitations}
\label{sec:quadsize}
For non-polar \px\ locations,  the $A_{2m}$ terms are non-zero at $m=1$ and/or 2 and produce acceleration components $\approx a_{\rm tidal} \cos m\phi$ around the tracer orbit.  These drive observable oscillating perturbations.  The quadrupole in particular is a distinctive signature of the tidal field.  Its characteristic amplitude, expressed as a physical or angular displacement is, for a tracer with period $P$,
\begin{align}
  \Delta x & \approx a_{\rm tidal} \Omega_0^2/4 =  \frac{\epsilon d_T}{4} \approx 6\,\textrm{m}\, \left(\frac{d_T}{5.2\,\textrm{AU}}\right)^4 \\
  \Delta \theta & \approx \Delta x/d_T =  \frac{\epsilon}{4} \approx 1.7\,\uas\, \left(\frac{d_T}{5.2\,\textrm{AU}}\right)^3.
\end{align}
These stronger $d_T$ dependences favor the use of the Trojans over the MBAs, and are of similar order to the collective accuracy expected from \lsst\ observations of Trojans, and much larger than spacecraft ranging accuracies.

Quadrupolar oscillations are also excited by less exotic gravitational sources, namely the quadrupole moments of the collective TNO and MBA populations, or the most massive individual bodies in these populations.  Main-belt quadrupoles should be distinguishable from \px\ quadrupoles by their different radial dependence across the giant-planet region ($r^{-3}$ vs $r^1$ force scaling) and the rapid orbital periods of the MBAs.  A TNO quadrupole is, however, very similar to one from \px, and must be limited either by prior constraint on its amplitude or by measuring some $\ell>2$ component that is known to be associated with the perturber's mass distribution.

From these rough calculations, we conclude that \px\ induces displacements that are potentially detectable at high significance through several channels,  justifying the more thorough numerical investigation of the available information.  \change{We also note that the signals from precession and from a shift in the period-radius relation continue to grow linearly with time, so the $S/N$ from a monitoring program operated continuously over time $t$ would improve as $t^{3/2}.$  The oscillatory quadrupole signal would gain $S/N$ more weakly, as $t^{1/2},$ over multiple orbital periods of the tracer.}

\section{Mathematical methods and model}
\label{sec:math}
We estimate the ability of a set of observational data points
$\textbf{d}=\{d_1,d_2,...,d_k\}$---in our case the measurements of
angular positions or ranges to solar system bodies---to constrain a set of parameters  $\textbf{p}=\{p_1,p_2,...,p_N\}$ of a model.  In our case the parameters are the masses of a set of active solar system bodies, and the phase space positions of active and tracer bodies at a reference epoch.  The parameters also include those governing various systematic effects in the measurements, such as differential chromatic refraction for ground-based astrometry.  We are interested in constraints on a single parameter---the mass of a putative PX---after marginalization over all the other ``nuisance'' parameters.  We will use the Cramér-Rao theorem~\citep{Cramer_1946,Rao}, which states that no unbiased estimator of the parameters can yield more precise constraints than calculated from the Fisher matrix, as described below.

\subsection{The Fisher matrix}

For a system with likelihood $\mathcal{L}(\textbf{d}|\textbf{p})$ of obtaining the data given the parameters, the available information on our parameter set is expressed by the $N\times N$ Fisher information matrix $\mathcal{F}$:
\begin{equation}
    \mathcal{F}_{ij} = \left<-\frac{\partial^2\log{\mathcal{L}}}{\partial p_i\partial p_j}\right>
\label{eqn:fisher}
\end{equation}

Let us consider an unbiased estimator $\theta=\{\theta_1,\theta_2,...,\theta_N\}$ of our parameter set $\textbf{p}$. The multivariate Cramér-Rao inequality imposes the following restriction on the covariance matrix of any such estimator:
\begin{equation}
    \text{Cov}(\theta) \geq \mathcal{F}^{-1}
\end{equation}
where the matrix inequality states that the difference matrix $\text{Cov}(\theta) - \mathcal{F}^{-1}$ is positive semidefinite (\eg\ $[\textbf{v}^T(\text{Cov}(\theta) - \mathcal{F}^{-1})\textbf{v}] \geq 0$ for any vector $\textbf{v} = \{v_1,v_2,...,v_N\}$). This condition allows us to establish a lower bound for the variance of any individual estimator $\theta_i$ of the parameter $p_i$. 

If the data points $\textbf{d}$ have independent gaussian errors $\sigma_{\textbf{d}} = \{\sigma_{1}, \sigma_{2}, ..., \sigma_{k}\}$ and means $\mu_{\textbf{d}} = \{\mu_{1}(\textbf{p}), \mu_{2}(\textbf{p}), ..., \mu_{N}(\textbf{p})\}$, the Fisher information becomes

\begin{equation}
       \mathcal{F}_{ij} = \sum_{l=0}^k\left<-\frac{\partial^2}{\partial p_i\partial p_j}\left(-\frac{(d_l-\mu_l(\textbf{p}))^2}{2\sigma_l^2}\right)\right>
\end{equation}
which simplifies to

\begin{equation}
\label{fisher_gaussian}
       \mathcal{F}_{ij} = \sum_{l=0}^k\frac{\partial \mu_l(\textbf{p})}{\partial p_i}\frac{\partial\mu_l(\textbf{p})}{\partial p_j}\frac{1}{\sigma_l^2}
\end{equation}

The lower bound for the unmarginalized variance of $\theta_i$ is
\begin{equation}
    \text{Var}(\theta_i) \geq (\mathcal{F}_{ii})^{-1}
\end{equation}
The process of marginalizing over all parameters $p_j$ for which $j\neq i$ weakens the derivable constraints $p_j$. Instead of simply inverting the $ii$ element, the whole matrix is inverted, and then the $ii$ element of the inverse Fisher matrix is selected. The bound becomes 
\begin{equation}
  \text{Var}(\theta_i) \geq (\mathcal{F}^{-1})_{ii}.
  \label{marginalization1}
  \end{equation}

Two additional properties of the Fisher matrix will be exploited here. First, the Fisher matrices of statistically independent experiments can be summed to express the information in their joint constraints.  For example, we can incorporate knowledge of a planet's mass from satellite tracking by adding $1/\sigma^2_M$ to the appropriate diagonal element of $\mathcal{F}$.
Second, we can generalize the equation for marginalization in \eqq{marginalization1} to the case where we
partition the parameters into a retained subset of interest $\textbf{p}_R$ and nuisance parameters subset $\textbf{p}_N.$ The marginalization over nuisance parameters transforms the Fisher matrix for the full parameter set into a smaller one over the retained parameters via
\begin{equation}
  \mathcal{F} = \left(\begin{array}{cc}
                       \mathcal{F}_{RR} & \mathcal{F}_{RN} \\
                       \mathcal{F}_{NR} & \mathcal{F}_{NN}
                     \end{array} \right)
                   \longrightarrow  \mathcal{F}_{RR}  - \mathcal{F}_{RN} \mathcal{F}_{NN}^{-1} \mathcal{F}_{NR} .
                   \label{marginalization2}
                 \end{equation}
                 
We construct a model for solar system dynamics and use it to compute the derivatives of solar system observables with respect to our model parameters. From those derivatives and from the uncertainties of the observational data points, we compute the Fisher matrix with \eqq{fisher_gaussian}.  
The inference of solar system parameters from these observations should closely match the Fisher bound, because (a) the uncertainties on the observations are generally well described by Gaussians, and (b) the dynamical model is very well constrained: any departures from a nominal model (i.e. with $M_{X}=0$) are in the regime where linear perturbation theory will suffice.  Under these conditions, the likelihood is formally Gaussian, and the Cramér-Rao bound is known to be saturated, \ie\ the attained uncertainty equals the lower bound.

The greater threat to the accuracy of Fisher forecasts is the possible omission from the model of physical or observational effects that are covariant with $M_X$ in the posterior likelihood.  We will aim to incorporate  any such effects into our model.

\subsection{Dynamical model and parameters}
\label{sec:model}
Our model for solar-system observations is a numerical integration of the Newtonian equations of motion, detailed in Section~\ref{sec:derivs}.  We are at liberty to ignore relativistic effects since they will not significantly alter the derivatives needed for the Fisher matrix in \eqq{fisher_gaussian}.  Only if the Fisher matrix is near singularity will small perturbations to the derivatives be important.  As long as we assume standard General Relativity (GR), there is no additional freedom to the model introduced by relativistic terms.  Likewise we can ignore the effects of gravitational lensing, stellar aberration, or other small but known perturbations to the observed positions.
  
The presence of a \px\ can be modeled through a free mass parameter $M_X$ at a fixed distance $d_X$ from the Solar System barycenter.  The effect of \px\ on the observable members of the solar system over the scale of a few decades can be decomposed into a constant acceleration (which is unobservable \change{because it does not create any differential motion between observers and targets}), a tidal acceleration scaling as $M_X/d_X^3$, and higher-order multipoles that are too small to be detectable with the data under consideration.  
A general tidal field is described by a traceless symmetric $3\times3$ tensor, or by the Hermitian coefficients of the $\ell=2$ spherical harmonics---either way, 5 real degrees of freedom.   It is convenient to introduce five nominal positions for \px, assigning to each a different mass parameter $M_{X1},\ldots,M_{X5}$, each located $d_X=400$~AU from the barycenter, at selected angular positions listed in Table~\ref{table:p5}.  Since any tidal field---including the tidal field of a \px\ at any position on the sky---can be expressed as a linear combination of the tidal fields generated by these 5 masses, the $5\times5$ covariance matrix of these five free parameters can be used to derive the expected uncertainty on any posited distant masses.  The mathematics for this operation are derived in Appendix~\ref{sec:tidal}. \change{Our choice of the 5 point-mass locations is not unique; any 5 locations for which the transformation matrix between point masses and quadrupole coefficients is well-conditioned would work.}

\begin{deluxetable}{cccc}
\tablecaption{Positions of canonical \px's
  \label{table:p5}
  }
\tablehead{
  \colhead{}  & \colhead{$X$} & \colhead{$Y$} & \colhead{$Z$}}
\startdata
 $\boldsymbol{\hat{n_1}}$ & 1 & 0 & 0 \\ 
 $\boldsymbol{\hat{n_2}}$ & 0 & 1 & 0 \\
 $\boldsymbol{\hat{n_3}}$ & $+\sqrt{3}^{-1}$ & $+\sqrt{3}^{-1}$ & $-\sqrt{3}^{-1}$ \\
 $\boldsymbol{\hat{n_4}}$ & $+\sqrt{3}^{-1}$ & $-\sqrt{3}^{-1}$ & $+\sqrt{3}^{-1}$ \\
 $\boldsymbol{\hat{n_5}}$ & $-\sqrt{3}^{-1}$ & $+\sqrt{3}^{-1}$ & $+\sqrt{3}^{-1}$
 \enddata
\tablecomments{The unit vectors (in ecliptic coordinates) pointing to the positions of the five fiducial \px's introduced to our dynamical model.  The tidal field of a perturber at any location in the sky can be expressed as a linear sum over these five, as detailed in Appendix~\ref{sec:tidal}.}
\end{deluxetable}

All other parameters of the model are ultimately considered nuisances in the pursuit of $M_X.$ The baseline model contains the following:
\begin{itemize}
\item For each of the eight known planets we consider seven free parameters: the six state vector components at the reference epoch $t_0$, taken to be MJD $=60000$ (25.0 Feb 2023), and the mass $M$, all given for the barycenter of the planet plus satellites.
The masses of the known planets (and of some main belt asteroids and TNOs) are constrained through close encounters with spacecraft or tracking of natural or artificial satellites. More precisely, it is $GM$ that is well constrained, but we will take $G$ as fixed and place all uncertainties in the masses, which is mathematically equivalent. We take this information into account by adding independent gaussian priors on each planet's mass to our Fisher matrix. The adopted priors are listed in Table~\ref{priors}. 

\item The Sun has a free mass $M_\odot$. 
There is a degeneracy in our Fisher matrix in that any overall shift or velocity of the solar system barycenter is unobservable.  Marginalization over the state vector of the barycenter is mathematically equivalent to treating the initial state vector of any single body as fixed.  We therefore choose to treat the state vector of the Sun as known.

\item We include the $J_2$ gravitational moment of the Sun as a free parameter. To establish an adequate prior, we consider the range of $J_2$ values derived by \cite{J2} through helioseismology. We then adopt $\sigma_{J2}=0.05$.
  
\item We ascribe a free mass parameter to each of the 343 largest main-belt asteroids. These are the same asteroids that are active in the JPL ephemerides DE430 and DE431 \citep{Folkner_2014}.  We fix their initial state vectors to the values in these ephemerides. For 20 asteroids which have precise masses determined from close encounters, we introduce individual Gaussian priors to the mass parameters. For Ceres and Vesta, the priors are listed in Table \ref{priors}. The other 18 asteroid priors are taken from the uncertainty of the weighted averages from \cite{Baer_2008}, from which we only leave out 189 Phthia, that is not included in our 343-asteroid set. For the remaining 323 asteroids, we assign a nominal mass based on their $H$ values, and introduce parameters for deviations from the nominal mass.
One general rescaling parameter multiplies all the masses simultaneously, and an additional parameter for each asteroid modulates its individual mass. The global parameter has a prior with standard deviation of 20\% from unity. The individual masses have independent priors with $\sigma$ at 50\% of their value. 

\item In addition to individually modeling the DE431 asteroids, we also model a set of two rings placed in the main belt to account for mass contributed by smaller asteroids.  These rings lie in the ecliptic plane with radii $2.06 \mathrm{AU}$ and $3.27 \mathrm{AU}$. Each ring's mass is a free parameter, to which Gaussian priors with $\sigma=3.3 \times 10^{-5}M_{\oplus}$ are assigned. The two radii mark the inner and outer bounds of the bulk of the main belt.  Using two rings introduces our uncertainty in the effective mass-weighted mean radius of the smaller bodies. \change{This value is chosen as a compromise between the different belt masses proposed in the literature \citep[\eg][]{Carry_2012,Pitjeva_2018}, an exact value is not needed as the derivatives are insensitive to the mass assumed.}

\item We include a free parameter for the mass of the Pluto system, leaving its state vector fixed due to its low mass. The adopted prior for this parameter is included in Table \ref{priors}.

\item We model the azimuthally symmetric portion of the Kuiper belt as the sum of two uniform rings of mass in the ecliptic plane at $39.5\,\mathrm{AU}$ and at $43.0\, \mathrm{AU}$ from the solar system barycenter, with distances chosen to correspond with the Plutinos at the 3:2 mean motion resonance with Neptune, and the classical belt's rough center.  Each is given a nominal mass of $10^{-2}M_\oplus$\change{, based on estimates by \citet{Pitjeva_2018}}. Since the true Kuiper belt mass is quite uncertain, we attribute to each mass a Gaussian prior with a $\sigma$ of 50\% of its value.  Again, the use of two rings introduces uncertainty in the effective radius of the TNOs. In particular this means that ratio of the $r^3$ and $r^4$  potentials that it induces on bodies interior to it is left free to vary.

\end{itemize}

\subsection{Degeneracies with trans-Neptunian quadrupoles}

Our assumed ring model for the Kuiper belt ignores the graininess and asymmetries in the TNO distribution. These features can generate stochastic tidal fields that are indistinguishable on decade timescales from a \px\ tidal field on observations interior to Neptune. In order to place priors on such quadrupoles and incorporate them into our results, we use the L7 synthetic debiased model from the CFEPS project \citep{Kavelaars_2009,Petit_2011,Gladman_2012} to recreate the evolution of TNO distribution features in the Neptune co-rotating frame over a time range of $2\times 10^4$ years. We sample the five components of the internal quadrupole field in 50 year intervals, and we measure the $5\times5$ covariance matrix of the quadrupole components from these samples. Variations are of the order of $10^{-3}M_{\text{CFEPS}}/(40\,\textrm{AU})^3$, where $M_{\text{CFEPS}}$ is the total mass of the population, estimated at $1.5 \times 10^{-2}M_\oplus$. The covariance matrix is then transformed into our basis of five point masses at 400 AU, and it is added to the covariance matrix obtained through the Fisher matrix method.

\subsection{Mass uncertainties from large TNOs}

The CFEPS sample reaches magnitudes as bright as $H\approx 3.4$.
We can reasonably suppose that all Kuiper Belt objects brighter than this limit are known or will be shortly.  Thus, the additional uncertainty in the internal gravitational field depends on how well their masses are determined. We add to the quadrupole covariance matrix the contribution of the 15 brightest TNOs, except Pluto which was modeled separately. (The potential existence of larger and more distant TNOs of gravitational significance can be considered the subject of this work.)  For Kuiper belt objects (KBOs) with mass determination from moons, we adopt the published system-mass uncertainties from such data.  For the others, we assign an uncertainty equal to half the mass estimated from their radius measurement and a presumed density. \change{These uncertainties are shown in Table \ref{priors}.}

Over periods of time much shorter than a KBO period, the quadrupole gravitational field generated by KBOs is essentially indistinguishable from a \px\ field from any observations of test bodies interior to Neptune.  This leads to an irreducible floor on the sensitivity of searches for gravity from \px. We will refer to the combined uncertainties from individual large-KBO masses, plus the stochastic tidal fields of smaller bodies described in the previous subsection, as the collective ``KBO floor'' on $\sigma_M$.
  The \change{contributions} of stochastic fields dominate this uncertainty, and are likely to continue to set a limit on detectability of \px\ tidal fields for many decades.

\begin{deluxetable}{|lcc|}
\tabletypesize{\small}
\tablewidth{0pt}
\tablecaption{Priors on $GM$ of planets}
\label{priors}
\tablehead{
    \colhead{Planet} & \colhead{$\sigma_{GM}/GM_\oplus$} & \colhead{Reference}}
\startdata
Mercury & $3.0 \times 10^{-9}$ & \cite{KONOPLIV2020113386}\\
Venus          & $1.5 \times 10^{-8}$ & \cite{KONOPLIV19993} \\
Earth\tablenotemark{a}   & $3.3 \times 10^{-8}$ & \cite{Pitjeva_2009}\\
Mars     & $1.3 \times 10^{-8}$ &  \cite{KONOPLIV2011401} \\
Jupiter & $2.1 \times 10^{-5}$ &  \cite{Durante_2020_JUNO} \\
Saturn & $2.8 \times 10^{-6}$ & \cite{Jacobson_2006}\\
Uranus & $1.1 \times 10^{-5}$ & \cite{Jacobson_2014}\\
Neptune & $2.5 \times 10^{-5}$ & \cite{Jacobson_2009} \\
Ceres & $9.9 \times 10^{-10}$ & \cite{Konopliv_Park_2018} \\
Vesta & $3.0 \times 10^{-11}$ & \cite{KONOPLIV2014103} \\
Pluto & $5.2\times 10^{-6}$& \cite{Stern2015}\\
Eris & $1.4 \times 10^{-5} $& \cite{HOLLER2021114130}\\
Makemake & $1.0 \times 10^{-5}$ & \textbf{presumed}\tablenotemark{b} \\
Haumea & $6.7\times 10^{-6}$ &\cite{Ragozzine2009}\\
Gonggong & $1.2\times 10^{-5}$ & \cite{KISS20193} \\
Quaoar &$1.5\times 10^{-5}$ &\cite{FRASER2013357}\tablenotemark{c} \\
Orcus & $3.2 \times 10^{-7}$ & \cite{GRUNDY201962} \\
Sedna & $8.6 \times 10^{-5}$& \cite{Pal2012}\tablenotemark{d} \\
2013 FY27 & $3.9 \times 10^{-5}$& \cite{Sheppard_2018}\tablenotemark{d} \\
Varda & $1.1 \times 10^{-6}$& \cite{GRUNDY2015130}\\
2014 UZ224 &$2.2\times 10^{-5}$& \cite{Gerdes_2017}\tablenotemark{d} \\
2002 AW197 & $4.0\times 10^{-5}$& \cite{vilenius2014}\tablenotemark{d}\\
G!kún$||$'hòmdímà & $5.5 \times 10^{-7}$& \cite{GRUNDY201930}\\
2002 TX300 &$2.1\times 10^{-6}$ & \cite{Elliot2010-zn} \\
Ixion & $3.1\times 10^{-5}$ & \cite{Levine2021}\tablenotemark{d} \\
    \enddata
    \tablenotetext{a}{Derived from constraints on the Earth-Moon mass ratio.}
        \tablenotetext{b}{We assume that future data on the Makemake system\change{, such as the tracking of its known moon MK2,} will constrain its total mass to a similar level of uncertainty as current constraints on Gonggong and Haumea.}
        \tablenotetext{c}{\citet{FRASER2013357} gives four possible orbit solutions for the Quaoar system. We assume this ambiguity will be solved during the following years.}
    \tablenotetext{d}{No dynamical mass measurement. Mass uncertainty assumed to be half the total mass estimate given its radius and $\rho \approx 2$ g/cm$^3$. Reference is for object radius.}
 \end{deluxetable}

\section{Numerical methods}
\label{sec:numerics}
\subsection{Determination of the derivatives}
\label{sec:derivs}

Variational derivatives for each body were determined by adding bodies and effects to a base simulation consisting of 9 massive bodies (the sun and 8 planets). Our simulation begins at MJD~60000.0, with the initial states and masses of the 9 massive bodies queried from the JPL Horizons service.\footnote{\url{https://ssd.jpl.nasa.gov/horizons/app.html}}

Beginning with this base simulation, we compute variational derivatives of planetary positions with respect to their $(8\times7)+1=57$ parameters using \texttt{REBOUND}'s variational equations module \citep{Rein_2016}. All simulations are integrated from the starting epoch for 13 years, using the 15th order adaptive \texttt{ias15} integrator \citet{Rein_2015,Everhart_1985} implemented in \texttt{REBOUND} \citep{REBOUND}. We choose \texttt{ias15} over symplectic integrators like \texttt{WHFast} since our integrated time span is relatively short (no longer than a few orbital periods of the Trojan asteroids).  A forward integration through 2036 covers all future observations considered in this paper, and a backward integration to 1965 covers historical data we use.

We add relevant bodies and forces to the system to calculate derivatives with respect to other sources of gravity.
First, we compute derivatives with respect to the \px\ parameters by adding a fiducial \px\ along each one of the five axes described  by the unit vectors in Table~\ref{table:p5},
with nominal mass of $M_X= 5 M_\oplus$ and circular orbit at $d_X=400\,\mathrm{AU}.$
Motion of \px\ during the integration has negligible effect, so the initial state vectors of the \px's are held fixed, leaving five free parameters to represent \px\ (equivalent to the five free parameters of a general tidal field). The nominal mass $M_X$ is chosen at the low end of the range suggested by
\citet{Brown_2021}.

Next we augment our base simulation by adding 343 massive main belt asteroids. We obtain the masses of these from Table~12 in \citet{Folkner_2014};\footnote{Note that there are 344 columns in table in \citep{Folkner_2014}, as the asteroid 499-Hamburga was listed twice.} these asteroids were used in the computation of the DE430 and DE431 JPL ephemerides. The states of these asteroids at the starting epoch were queried from JPL Horizons.  The size of these simulations (more than $350$ massive bodies, in addition to several thousand test particles once we examine the Trojan asteroids in Section~\ref{sec:trojanresults}) necessitates parallelization of the integration.  We perform this task by dividing the Trojan asteroids and massive main belt asteroids into $10$ and $30$ groups respectively. We then compute variational derivatives for each combination of every subgroup of main belt asteroids and Trojans, at all \px\ positions, and compiled the results. Since each individual simulation only contains a subset of massive asteroids, this means that the trajectories of the bodies in each subgroup are slightly inconsistent with each other. Fortunately, the main belt asteroids contribute only a small perturbation to the base simulation of Trojans and planets\change{. The addition of a typical single asteroid causes a relative change in the variational derivative of another asteroid of $\leq 10^{-8}$, so} this small inconsistency can be neglected for the purposes of making Fisher-matrix derivatives. 


The potential from each ring \change{modeled in Section~\ref{sec:model}} felt by a particle of mass $m_p$ at radius $r_x$ can be written as 
\begin{equation}
    \phi = - \frac{G M }{2 \pi} \int_{0}^{2 \pi} \frac{1}{r_0^2 + r_x^2 + z^2 - 2 r_0 r_x \cos{\theta_0} } d\theta_0.
\end{equation}
where $r_0$ is the radius of the ring and $M$ is its mass. The force due to this potential is added to the integrator using the \texttt{REBOUNDx} package \citep{REBOUNDx} and the derivatives are computed by finite differencing two simulations with slightly different ring masses. 

Finally, we compute derivatives with respect to the $J_2$ of the Sun again by finite differencing two simulations with a slightly different value of $J_2$ about an initial value taken of $2.2 \cdot 10^{-7}$ \citep[\eg][]{Pijpers_1998}. The Sun's north pole is tilted from the north ecliptic pole by $16.12 ^{\circ}$ in right ascension and  $-2.70 ^{\circ}$ in declination, which can be computed using coordinates of the solar north pole \citep{Archinal_2018} and  the obliquity of the earth  \citep[\eg][]{Capitaine_2003}.

In the solar frame, where the $\hat{z}$ axis aligns with the solar pole, the force from a $J_2$ can be written as

\begin{equation}
    \vec{F} = \frac{3 G M_\odot J_2 R_\odot}{2 r_p^5} \left[\left(\frac{5 z_p^2}{r_p^2} - 1 \right) (x_p \hat{x} + y_p \hat{y}) + \left(\frac{5 z_p^2}{r_p^2} - 3 \right) (z_p \hat{z}) \right]. 
\end{equation}
where $x_p$,$y_p$,$z_p$ are the cartesian distances from the sun along each axis, and $r_p = \sqrt{x_p^2 + y_p^2 + z_p^2}.$
We add this force (after rotating into the ecliptic plane) into \texttt{REBOUND} using the \texttt{REBOUNDx} package \citep{REBOUNDx}.  Only a single free parameter is involved, since the orientation of the Sun's rotation pole is known to within about $0.1^\circ$ \citep{Beck_2005}.

For a given observer our Cartesian positions can be converted to an observable position on the sky through $\alpha = \arctan(y(t)/x(t))$, $\delta = \arcsin(z(t)/\sqrt{x(t)^2 + y(t)^2 + z(t)^2})$, where $\alpha$ and $\delta$ are the right ascension and declination of the body and $x(t)$, $y(t)$ and $z(t)$ are all Cartesian differences between the coordinates of the body in question at the time the signal is emitted and the coordinates of the observatory at the time of observation (\eg\ $x(t) =  x_p(t - \tau) - x_{obs}(t)$, for some light-time delay $\tau$). Conversion between our computed Cartesian state derivatives and the astrometric derivatives can be computed with the Jacobian. From the cartesian state positions and its derivatives, we also compute derivatives of round-trip light time observables, in order to incorporate spacecraft ranging into our Fisher analysis.

\startlongtable
\begin{deluxetable}{lcccc}
\tabletypesize{\small}
\tablewidth{0pt}
\tablecaption{Planetary observations assumed for fitting}
\label{historical_data}
\tablehead{
    \colhead{Planet} & \colhead{Observatory/Spacecraft\tablenotemark{a}} & \colhead{Type\tablenotemark{b}} & \colhead{Time Span} & \colhead{Typical uncertainty $/\sqrt{N}$}\tablenotemark{c}}
\startdata
Mercury & Messenger & A & 2008--2009 & 0.5~mas \\
             & Messenger & R & 2008--2009 & 20~m \\
             & Messenger & R & 2011--2015 & 0.02~m \\
Venus & Magellan/VLBI & A & 1990--1994 & 0.4~mas\\
 & Cassini & A & 1998--1999 & 3~mas\\
 & Cassini & R & 1998--1999 & 3~m \\
 & Venus Express/VLBI & A & 2007--2013 & 0.3~mas \\
 Mars  & Mariner 9 & R & 1971--1972 & 3~m \\
         & Viking & R & 1976--1982 & 0.5~m \\
         & Phobos 2 /VLBI & A & 1989 & 3~mas \\
         & Pathfinder & R & 1997 & 4~m\\
         & MGS & R & 1999--2006 & 0.002~m \\
         & MGS/VLBI & A & 2001--2003 & 0.3~mas \\
         & Odyssey         & R & 2002--2013 & 0.008~m \\
         & Odyssey/VLBI & A & 2002--2013 & 0.03~mas\\
         & MRO & R & 2006--2014 & 0.02~m \\
         & MRO/VLBI & A & 2006--2013 & 0.03~mas\\
         & Odyssey \textbf{(presumed)} & R & 2014--2034 & 0.006~m\\
         & MRO \textbf{(presumed)}& R & 2014--2034 & 0.009~m\\
 Jupiter & Pioneer 10/VLBI & A & 1973 & 100~mas\\
         & Pioneer 10 & R & 1973 & 4000000~m\\
         & Pioneer 11/VLBI & A & 1974 & 30~mas\\
         & Pioneer 11 & R & 1974 & 3000~m\\
         & Voyager 1/VLBI & A & 1979 & 30~mas\\
         & Voyager 1 & R & 1979 & 3000~m\\
         & Voyager 2/VLBI & A & 1979 & 3~mas\\
         & Voyager 2 & R & 1979 & 3000~m\\
         & Ulysses/VLBI & A & 1992 & 5~mas\\
         & Ulysses & R & 1992 & 20~m\\
         & Galileo/VLBI & A & 1996--1997 & 2~mas \\
         & Cassini & A & 2000 & 80~mas\\
         & Cassini & R & 2000 & 4000~m\\
         & Juno & R & 2016--2017 & 7~m\\
         & Juno \textbf{(presumed)} & R & 2018--2025 & 3~m\\
Saturn & Pioneer 11 & R & 1979 & 2000000~m\\
           & Pioneer 11 & A & 1979 & 600~mas\\
         & Voyager 1 & A & 1980 & 100~mas\\
         & Voyager 1 & R & 1980 & 50~m\\
         & Voyager 2 & A & 1981 & 200~mas\\
         & Voyager 2 & R & 1981 & 100~m\\
         & Cassini/VLBI & A & 2004--2011 & 0.6~mas \\
         & Cassini & R & 2004--2017 & 0.2~m\\
Uranus & Voyager 2 & A & 1986 & 10~mas\\
          & Voyager 2 & R & 1986 & 200~m\\
Neptune & Voyager 2 & A & 1989 & 4~mas\\
          & Voyager 2 & R & 1989 & 30~m\\
    Outer  & Nikolaev Obs. & A & 1966--1998 & 4~mas\\
    Outer  & La Palma Obs. & A & 1984--1998 & 3~mas\\
    Outer  & Bordeaux Obs. & A & 1985--1996 & 10~mas\\
    Outer  & Flagstaff Obs. & A & 1995--2015 & 0.9~mas\\
    Outer  & Table Mountain Obs. & A & 1997--2013 &1~mas \\
    \enddata
    \tablenotetext{a}{MGS: Mars Global Surveyor; MRO: Mars Reconnaissance Orbiter}
    \tablenotetext{b}{A: astrometry; R: range}
    \tablenotetext{c}{\change{N: number of data points.} For the astrometric data with different errors on ra and dec, uncertainty listed is the largest one.}
 \end{deluxetable}

\section{Results from major-planet tracking}
\label{sec:results}
In this section we forecast the \px\ constraints available from the set of published historical observations, with some extension to presumed future spacecraft data. In Section~\ref{sec:trojanresults}, we will consider the addition of a set of predicted LSST observations of 7664 Trojan asteroids.

\subsection{\change{Assumed observations}}
\label{sec:observations1}

The available historical data consists of astrometric measurements, which give the angular position of the planet or one of its moons, and radar ranging measurements, which give the light time between the observing station on Earth and a spacecraft at the planet (typically referenced to the planetary barycenter). The observations we use are presented in Table \ref{historical_data}, and are comprised of similar data to those compiled by \citet{Park_2021}. We restrict ourselves, however, to the data currently available on the JPL website\footnote{\url{https://ssd.jpl.nasa.gov/planets/obs\_data.html}}, plus the complete set of \textit{Cassini} data points available on the INPOP Astrometric Planetary Database website\footnote{\url{http://www.geoazur.fr/astrogeo/?href=observations/base}}. 

We posit that Juno observations of Jupiter's range will continue up to 2025, simulating data points with the same errors and spacing in time as the actual Juno data from 2016 and 2017. This corresponds to an addition of $21$ data points with spacing of $4-5$ months. For Mars we assume that two probes will be operating at all times between 2014 and the end of LSST observations, taking range measurements with similar spacing and precision as those from Odyssey and Mars Reconnaissance Orbiter.  These observations are marked as \textbf{presumed} in Table~\ref{historical_data}. The extension adds a total of $5943$ data points for MRO  and $16065$ for Odyssey. Both numbers are found by multiplying the number of data points taken in $2013$ by $21$ years.

\change{The assumed priors on masses of individual major and minor planets, summarized in Table~\ref{priors}, are independent of the positional data in Table~\ref{historical_data}, since they are derived solely from tracking of natural or artificial satellites around each mass.}

\subsection{Nominal Case}

\begin{figure}
  \centering
  \includegraphics[width=0.7\textwidth]{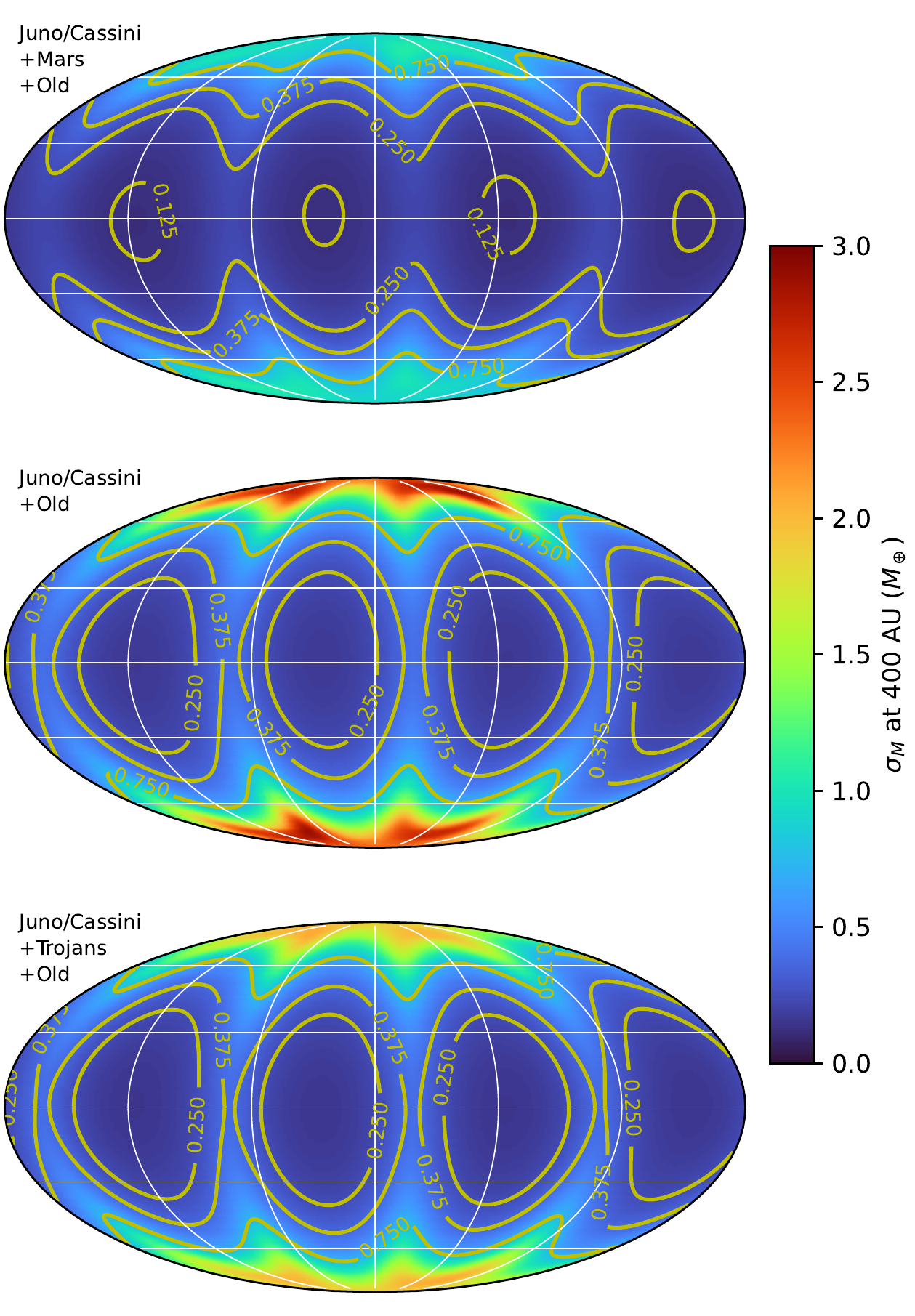}
  \caption{Maps of forecasted uncertainty $\sigma_M$ on the mass of Planet X at 400~AU as a function of its location in the sky.  Plots are in ecliptic coordinates with the Vernal equinox at center. \change{The Mollweide projection is used.} The upper panel shows the \emph{nominal} case using all observational constraints to date, including Juno/Cassini ranging and Mars ranging.  ``Old'' data are the existing measurements without Juno, Cassini, or Mars ranging. In the middle panel, only the old data and the Juno/Cassini data are used (Mars ranging is omitted)\change{, as described in Section~\ref{sec:subsets}}.  In the lower panel, forecasted \lsst\ observations of Jovian Trojans are included, which makes up some of the precision lost by omission of Mars \change{(see Section~\ref{sec:combinations}}).}
\label{maps}
\end{figure}

We start with our ``nominal'' case, where we use the full set of published historical planetary data, augmented by our presumed data for Mars and Jupiter only up to 2021. This simulates data which should already be in hand. The top panel of Figure~\ref{maps} plots the measurement noise $\sigma_M$ of a \px\ at 400~AU vs position in the sky for this nominal case.

For each observational scenario we consider, we construct such an all-sky map, and from it extract several summary statistics given in Table~\ref{limitstab}.  We list the minimum, median, and maximum values of $\sigma_M$ across the sky.  We also tabulate the percentage of the sky $f_{400}$ for which $\sigma_M<1,$ meaning that the presence of a $5M_\oplus$ \px\ at a distance of 400~AU would be detected unambiguously, at $5\sigma$ significance.  This is 99.2\% for the nominal case, \ie\ a \px\ at the near end of its expected range would be found anywhere on the sky.  A more stringent criterion, demanding $5\sigma$ detection of $5M_\oplus$ at a distance $d_X=800$~AU ($\sigma_M<0.125 M_\oplus$ on our plots) is satisfied over $f_{800}=4.8\%$ of the sky in the nominal case.

\begin{deluxetable}{l|ccc|rrr|rr}
  \tabletypesize{\small}
  \tablecaption{Limits on Planet X mass in multiple scenarios}
\label{limitstab}
\tablehead{\colhead{Comment} & \multicolumn{3}{c}{Data included}
  & \multicolumn{3}{c}{$\sigma_M$ for Planet X ($M_\oplus$)}
  & \multicolumn{2}{c}{$f_{\rm sky}$ detecting $5M_\oplus$\tablenotemark{a}} \\
   & \colhead{J/C\tablenotemark{b}} & \colhead{Mars} & \colhead{Trojans} & \colhead{Min} & \colhead{Median} & \colhead{Max} & \colhead{400 AU} & \colhead{800 AU}}
\startdata
Nominal & $\times$ & $\times$ &  & 0.12 & 0.22 & 1.07 & 99.2\% & 4.8\% \\
...published ranging & $\times$ & $\times$ &  & 0.12 & 0.24 & 1.27 & 94.0\% & 1.3\% \\
...ranging through 2034 & $\times$ & $\times$ &  & 0.11 & 0.20 & 0.80 & 100.0\% & 8.1\% \\
Old only &  &  &  & 9.11 & 12.38 & 43.41 & 0.0\% & 0.0\% \\
...$+$Juno/Cassini & $\times$ &  &  & 0.16 & 0.35 & 2.87 & 87.3\% & 0.0\% \\
...$+$Mars &  & $\times$ &  & 1.16 & 1.54 & 4.23 & 0.0\% & 0.0\% \\
Nominal$+$Trojans & $\times$ & $\times$ & $\times$ & 0.12 & 0.22 & 0.98 & 100.0\% & 5.0\% \\
...Trojan errors$\div 5$ & $\times$ & $\times$ & $\times$ & 0.11 & 0.20 & 0.58 & 100.0\% & 7.3\% \\
...Trojan errors$\div 20$ & $\times$ & $\times$ & $\times$ & 0.11 & 0.14 & 0.27 & 100.0\% & 29.2\% \\
Old$+$Trojans, no sys &  &  & $\times$ & 1.66 & 1.90 & 3.40 & 0.0\% & 0.0\% \\
 Old$+$Trojans &  &  & $\times$ & 1.84 & 2.15 & 3.88 & 0.0\% & 0.0\% \\
...$+$Juno/Cassini & $\times$ &  & $\times$ & 0.15 & 0.33 & 2.07 & 89.2\% & 0.0\% \\
...$+$Mars &  & $\times$ & $\times$ & 1.02 & 1.22 & 2.39 & 0.0\% & 0.0\% \\
KBO floor\tablenotemark{c} &\nodata  &\nodata  &\nodata  & 0.053 & 0.066 & 0.071 & 100.0\% & 100.0\%
\enddata
\tablenotetext{a}{Fraction of the sky over which a $5M_\oplus$ Planet X would be a $\ge5\sigma$ detection.}
\tablenotetext{b}{Ranging data from Juno and Cassini.}
\tablenotetext{c}{Limits purely from degeneracies with uncertainties in Kuiper Belt quadrupole.}
\end{deluxetable}

\subsubsection{Extended ranging}
\begin{figure}
  \centering
  \includegraphics[width=0.5\textwidth]{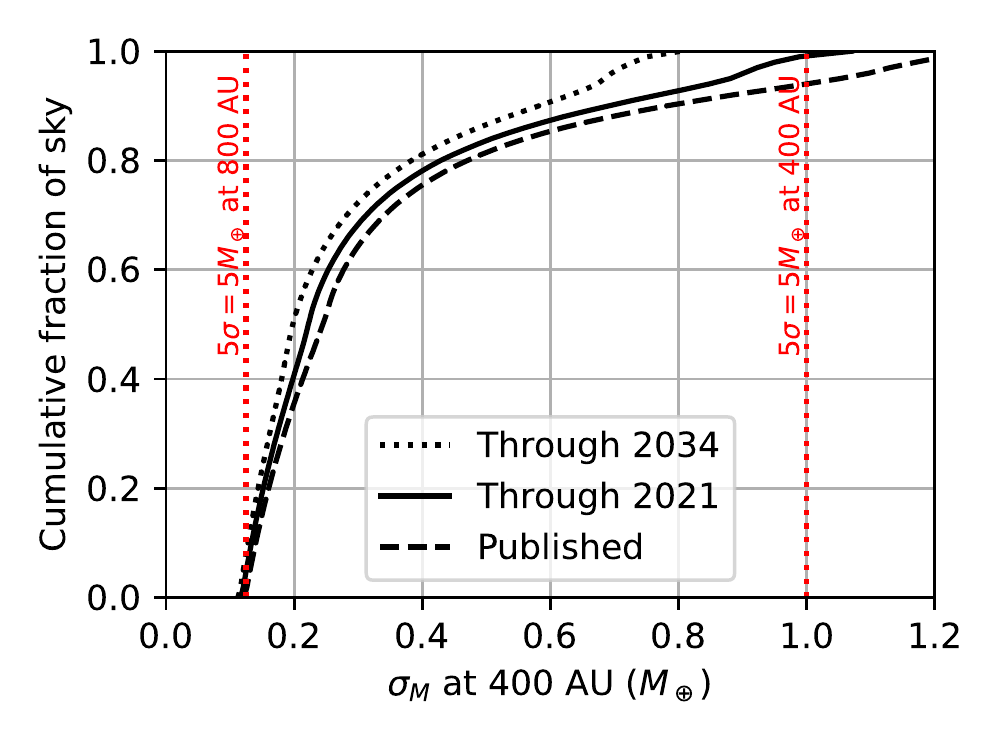}
  \caption{Fraction of the sky for which the $\sigma_M$ of a Planet X at 400~AU is below the value on the $x$ axis.  This is shown for the nominal scenario (solid), one restricted to already-published ranging, and another extrapolating ranging data through 2034. }
  \label{fig:fosky}
\end{figure}

Figure~\ref{fig:fosky} shows the effect of restricting the  Mars, Jupiter and Saturn ranging data to observations already published (dotted line), or extending the ranging with all presumed data up to 2025 for Juno, and up to 2034 for two Mars probes. We compare, for each, scenario, the fraction of the sky for which each level of detection is possible. Progressing from published$\rightarrow$current$\rightarrow$predicted data adds 5--15\% of sky coverage attaining any particular $\sigma_M.$

\subsubsection{Degraded ranging}

\begin{figure}
  \centering
  \includegraphics[width=0.5\textwidth]{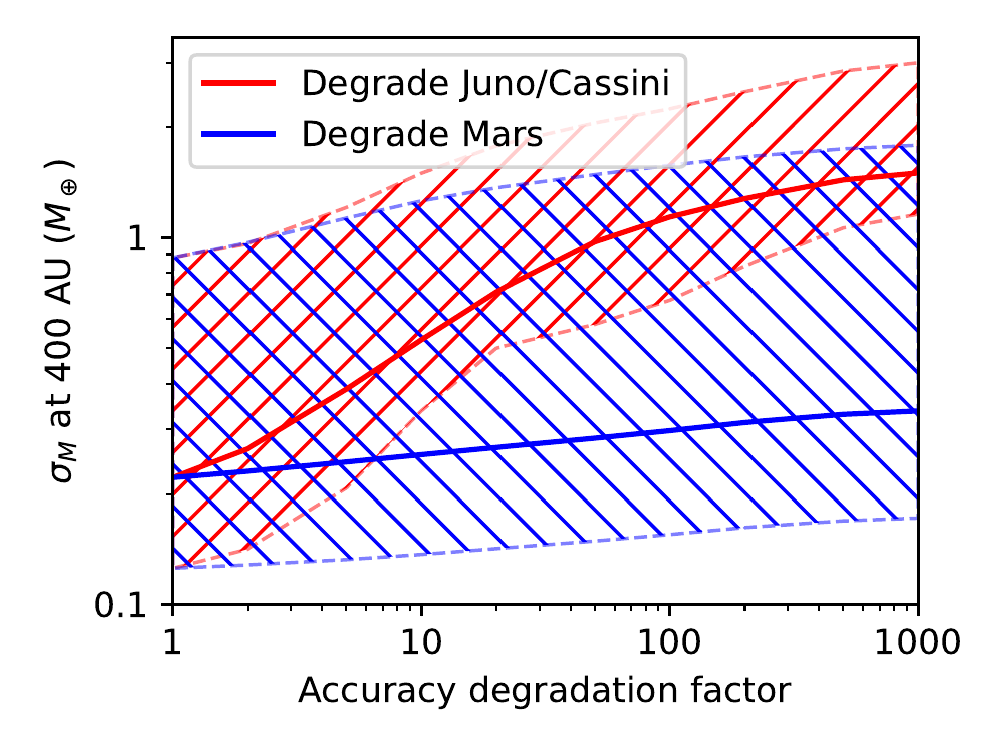}
  \caption{As we increase the measurement errors on the Juno/Cassini ranging (red) or Mars ranging (blue), the forecasted uncertainties $\sigma_M$  on the \px\ mass at 400~AU are degraded. The solid lines show the median value of $\sigma_M$ and the hatched regions bound the 5--95\% percentiles of $\sigma_M$ across the sky.  The full power of the J/C data are more critical to the \px\ constraints than that of the Mars ranging.
  }
  \label{fig:degrade}
\end{figure}

We also consider the scenario in which the uncertainties of the most precise ranging data are currently underestimated due to systematic errors.  In Figure~\ref{fig:degrade} we plot the ranges of $\sigma_M$ constraints as we inflate the \change{measurement} errors on either the Juno/Cassini ranging or on the Mars ranging.
The $\sigma_M$ constraints are seen to be only very weakly dependent on the accuracy of the Mars ranging, but are more sensitive to the Juno/Cassini uncertainties.

\subsection{Data subsets}
\label{sec:subsets}
To determine the relative value of different planets' data to the \px\ constraints, and to explore whether complementary sets of data could confirm each other's detections, we split the nominal data into three distinct subsets: First, the Juno$+$Cassini ranging data; second, the Mars ranging data;  and third, the remaining elements of the compilation summarized in Table~\ref{historical_data}. We will refer to this last subset as the ``old'' observations (even though some of the Mercury and Venus data are recent).

\begin{figure}
  \centering
  \includegraphics[width=0.9\textwidth]{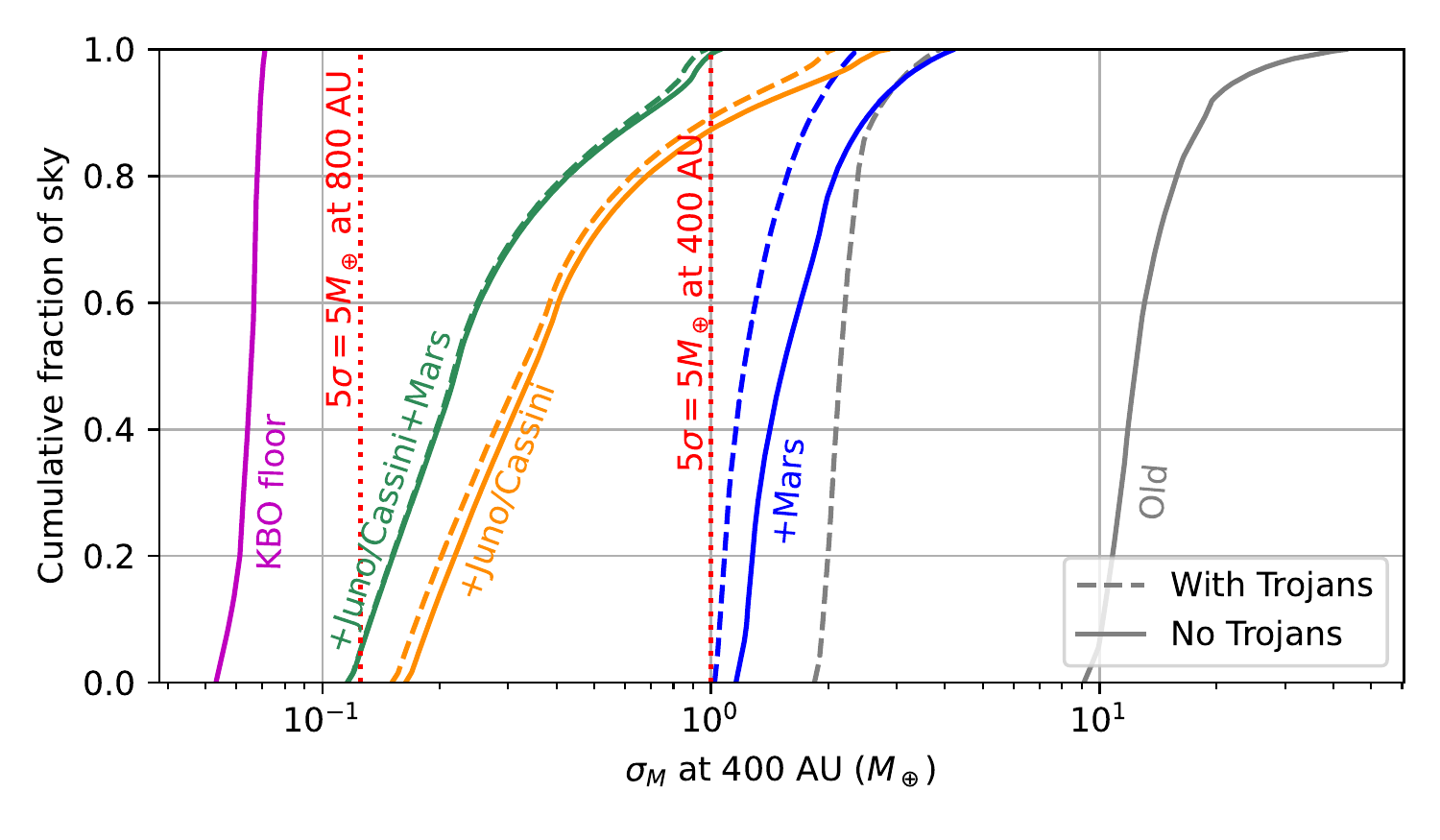}
  \caption{Each curve plots the fraction of the sky ($y$ axis) for which the uncertainty $\sigma_M$ on \px\ at 400~AU is at or below the value on the $x$ axis.  The solid curves are for cases without \lsst\ Trojan data, and are labelled by the subsets of the observational data they use.  All curves make use of the ``old'' observations. The dashed curves add the use of \lsst\ Trojan observations to the scenario of the solid curve of matching color.  The leftmost ``KBO floor'' curve plots the constraints that would be attained if the only limitation were confusion with the uncertain tidal field produced by the members of the Kuiper belt.  This floor is up to an order of magnitude below constraints with the best data expected in the next decade. The two vertical lines mark the $\sigma_M$ values that are needed for a definitive $5\sigma$ detection of \px\ if it has mass of $5M_\oplus$ and is at distance of 800~AU (left) or 400~AU (right).  Ideally we would be able to detect the former at 100\% of locations on the sky.}
\label{skyfrac}
\end{figure}

We find that the quality of $\sigma_M$ constraints improves as we move from Old, to Old$+$Mars, to Old$+$J/C, to the full Old$+$Mars$+$J/C nominal data.  This is illustrated by the solid curves in Figure~\ref{skyfrac}, which plots the fraction of the full sky achieving a given level of $\sigma_M$ in each scenario.  Numerical data on these scenarios are also given in Table~\ref{limitstab}.  Here it is clear that the J/C data are significantly more useful than the Mars data, however their combination is needed if we wish to approach secure detections of the most pessimistic $5M_\oplus,$ $d_X=800$~AU model across a substantial portion of the sky.

The middle panel of Figure~\ref{maps} shows the sky map of $\sigma_M$ in the Old$+$J/C case, illustrating the weaker (but still interesting) level of constraint compared to the upper, nominal case.

\subsection{Kuiper Belt quadrupole limit}

\begin{figure}
 \centering
  \includegraphics[width=0.6\textwidth]{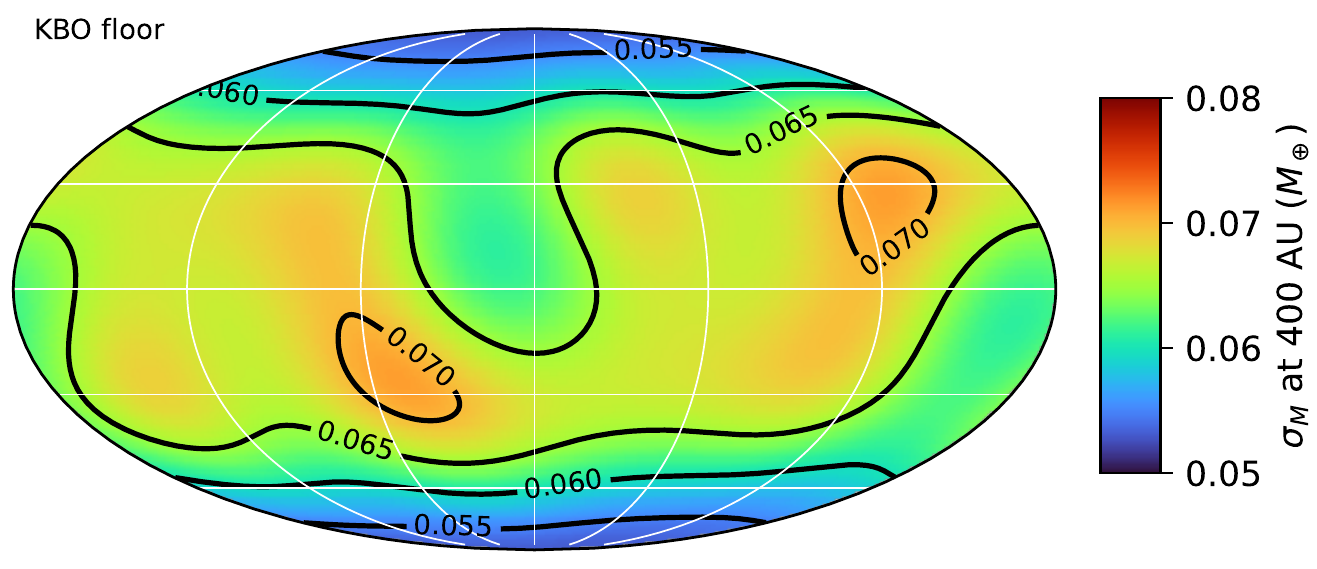}
  \caption{A map of $\sigma_M,$ the lower limit on achievable $M_X$ uncertainty for \px\ at 400~AU, set by ignorance of the Kuiper Belt's tidal field.  Note the change in color scale from Figure~\ref{maps}.}
  \label{fig:kbo}
\end{figure}

The portion of the Kuiper Belt which was treated separately from the Fisher matrix, that is, the uncertainty on the tidal field caused by asymmetries in population distribution and specific point sources whose mass are uncertain, amount to a fundamental limit, computed as an additive contribution to the covariance matrix. If this additive term is considered by itself, we achieve the results labeled as ``KBO floor'' in Figure~\ref{skyfrac} and in Table~\ref{limitstab}.   Figure~\ref{fig:kbo} is a sky map of the $M_X$ uncertainties at 400~AU induced purely by our ignorance of the Kuiper Belt mass distribution.  Note that a polar \px\ is better constrained in this case than one in the ecliptic plane, unlike our other cases where the pole is the weak spot.

\section{Trojans as test particles}
\label{sec:trojans}
In this and following sections we investigate the value of \lsst\ tracking of Jovian Trojans as constraints on \px.   A similar analysis could incorporate other minor body populations, especially the main belt (MBAs), but we limit our current scope to the Trojans. Future work could also extend this analysis to gauge the utility of \lsst-scale minor planet precision astrometry for other gravitational signals, \eg\  determination of masses of members of these populations.

\subsection{Measurement uncertainties}
\subsubsection{Astrometric precision per observation}
Our model for uncertainties in individual astrometric measurements of minor planets by \lsst\ assumes optimal extraction in the face of  three sources of random error.  Faint sources are dominated by uniform shot noise from diffuse foregrounds and backgrounds. Brighter sources become dominated by the shot noise from their own photons.  The brightest sources are limited by uncorrected stochastic atmospheric deflections, which we will place at a nominal level $\sigma_{\rm min}=2$~mas per axis.

The LSST survey simulation \citep{cadence_2022} predicts a set of characteristics of each observation of a run-through of the entire survey.  We make use of the tabulated values of the sky brightness (converted to counts per exposure per unit sky area $b$), the equivalent Gaussian width $\sigma_{\rm psf}$ of the point-spread function (PSF), and the filter band $Q \in \{g,r,i,z\}$ of each exposure (we ignore the $u$ and $Y$ bands as unlikely to add useful constraints).  Estimated values of the magnitude zeropoints $m_{Q,0}$ for each filter---\ie\ the magnitude generating a single photocarrier during the exposure---are taken from \citet{LSSTWebsite}.  We simplify the survey simulator's weather predictions by assuming that all nights with $<1\%$ cloud cover have no cloud extinction, and discarding observations with higher cloud cover.

The assumed absolute magnitude $H$ of the target is combined with the its geocentric and heliocentric distances $d_g$ and $d_h$ to yield a true apparent magnitude
in filter $Q$ of
\begin{equation}
    m_Q = H + 5\log_{10}{(d_gd_h)} + c_Q
\end{equation}
where $c_Q$ is the typical $Q-V$ color of the population. We have ignored the illumination-phase term, since the phase angle never exceeds $\sin^{-1} \frac{1\,\textrm{AU}}{d} \approx 11^\circ$ for Jovian Trojans, and will not significantly impact the overall $M_X$ sensitivity.
The photon count is then given by
\begin{equation}
    f = 10^{0.4(m_{Q,0}-m_Q)}.
\end{equation}

The best achievable astrometric uncertainty for an unresolved source is derived using Fisher matrix algebra in Appendix~\ref{sec:astrometry}.  Reproducing \eqq{eq:sigmax}, a circular Gaussian PSF with dispersion $\sigma_{\rm PSF}$ on each axis yields an astrometric uncertainty on each axis of:
\begin{equation}
    \sigma_x^{-2} = \frac{f}{\sigma_{\rm{PSF}}^2}\int_0^\infty u\frac{e^{-2u}}{e^{-u}+ \nu}du
\label{noisesigma}
\end{equation}
where
$\nu = \frac{2\pi\sigma_{\rm{PSF}}^2 b}{f}.$
Experience with the \textit{Dark Energy Survey (DES)} suggests that PSF-fitting astrometry can closely approach these theoretical limits. 

The uncalibrated portion $\sigma_{\rm min}$ of the atmospheric turbulence displacements then must be added in quadrature to the $\sigma_x$ value of \eqq{noisesigma} to yield the final $\sigma_x$ of the observation.  
\citet{Fortino_2021} demonstrate that calibration of \textit{DES} images based on the position of \textit{Gaia} stars can bring this error to $\sigma_{\rm min}\approx 2$~mas.  We consider this as our nominal value, though we also evaluate the effect that this value has on our final results in Section~\ref{sec:varySigmaMin}.

\subsubsection{Trojan catalog}
The list of all known Jovian Trojans was downloaded from the Minor Planet Center database\footnote{\change{Date of download: 07-Oct-2021.}} \citep{mpc_2021}, giving orbital elements and $H$ values for each.  The known sources exhibit a power-law size distribution, \ie\ exponential in $H$, for $H\le 14.5$, and we presume the observations are nearly complete to this magnitude.   For $H>14.5$, where the known counts depart from the exponential, we generate artificial Trojans by replicating the orbital elements of randomly selected known Trojans and assigning them new $H$ values until we match the extrapolation of the exponential distribution.

This extrapolation ends up being irrelevant for the final $M_X$ uncertainties because, as we will show below, most of the information is carried by Trojans in the regime where the observational catalog is nearly complete.

\subsubsection{LSST Observing Plan}
\label{sec:lssterrs}
We use the Baseline 2.0 simulated 10-year \lsst\ survey produced by the Survey Cadence Optimization Committee \citep{cadence_2022} which lists the times, pointings, and observing conditions of every exposure during the simulated survey.  The effects of varying airmass are reflected in the values of expected sky brightness and PSF size in the table, so the simulation is realistic in these regards, and the simulation also includes cloud conditions based on weather history at the site. \change{The total amount of cloudless exposures with $Q \in \{g,r,i,z\}$ is $\sim 1.2\times 10^6$.}

Because the orbital period of the Trojans is 12~years, we assume with modest optimism that \lsst\ tracking of Trojans will continue for 2 years past the nominal 10-year \lsst\ duration.  This also makes the results insensitive to the actual start date of the survey.  To implement this, we duplicate the pointings and weather from years 9 and 10 to create years 11 and 12 of our nominal survey.

At this point we have a catalog of all of the useful observations $j$ of Trojan target $i$ over the full 12-year survey, with the times of each and the per-axis measurement error $\sigma_{ij}$ for the object's position.  We will assume that these errors are statistically independent, which will be true for errors from shot noise and atmospheric turbulence.

\subsubsection{Information vs depth}
A useful measure of the constraining power of observations of some minor body population is the collective uncertainty $\sigma_{\rm pop}$ defined through this sum over the astrometric uncertainties $\sigma_{ij}$ per axis in observations $j$ of bodies $i$  with absolute magnitude $H_i$ brighter than some threshold:
\begin{equation}
  \sigma_{\rm pop}^{-2}(<H) = \sum_{i: H_i<H} \sum_j \sigma^{-2}_{ij}.
\end{equation}
This quantity gives the RMS size of a perturbation that can be detected (per axis) at $1\sigma$ using astrometry of the full population, in the absence of systematic measurement errors, or other free parameters that are covariant with the perturbation.

\begin{figure}
\begin{center}
  \includegraphics[width=0.5\columnwidth]{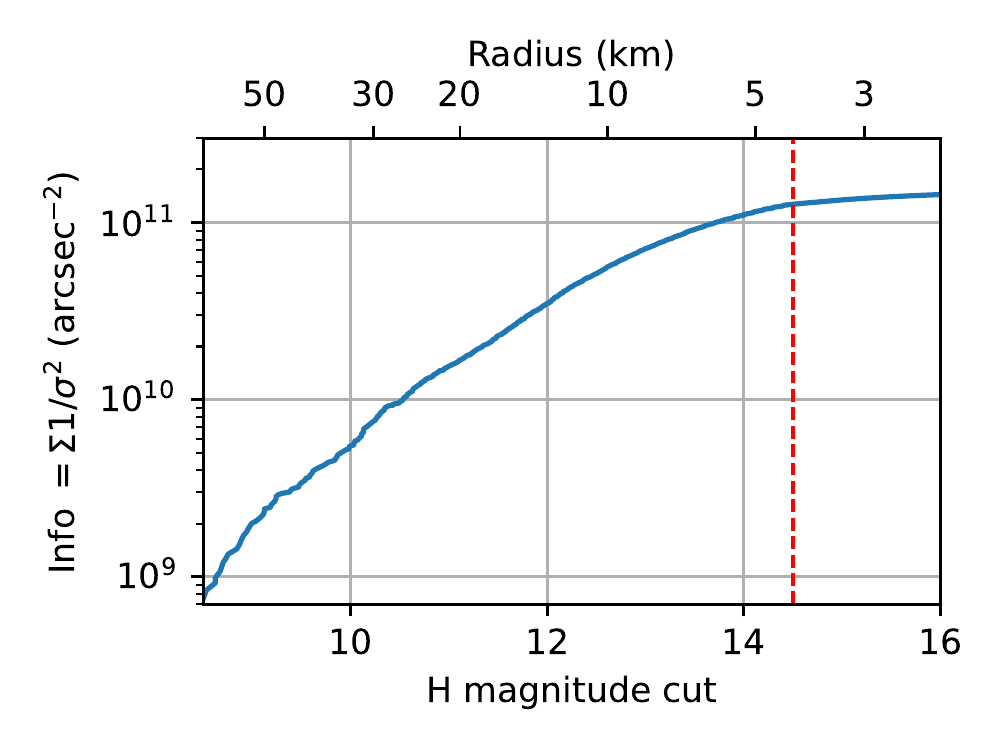}
\end{center}
\caption{Cumulative information from predicted LSST observations of the Trojans as a function of the magnitude of the faintest Trojan considered. Top axis shows object radius of objects with albedo of 0.04.  Trojans fainter than $H=14.5$ (vertical dashed line) do not add significant information---this is roughly the point at which their astrometry becomes limited by background noise.}
  \label{fig:sigpop}
\end{figure}

Figure~\ref{fig:sigpop} shows this quantity vs the limiting magnitude $H$ for the Trojan population.  This sum asymptotes to $\approx 3\,\uas$ at $H<14.5$ for the Trojans.  For the remainder of our analysis, we will restrict our analysis to the real \change{Trojan} sample (with $H<14.5$) leaving 7664 Trojans, which will be treated as massless test particles in further dynamical analyses.

Taking nominal distances and albedos of 5.2~AU and 0.04 for Trojans, the lower limit to radius $R_T$ for Trojans that would contribute significantly to the full-population constraints is roughly 4~km.
An error of $\sigma_{\rm min}=2$~mas subtends $\approx8$~km, twice the minimum radii of targets of interest.  At the faint end of the useful $H$ range, however, shot noise is dominant over atmospheric turbulence, so the per-visit centroiding errors are larger than the angular sizes of the objects. For brighter Trojan targets, the measurement errors per epoch are smaller than their radii. In either case, any source of \emph{stochastic} noise (uncorrelated between observations) becomes unimportant if it is well below the $\sigma_{ij}$ value.
At ground-based seeing of 0\farcs5--1\farcs0, all of the minor-planet targets will be unresolved, and trailing in $<30$~s \lsst\ exposures will negligibly degrade the astrometric accuracy, so we can assume point-source values for measurement errors.

The $\sigma_{\rm pop}$ errors are far below the sizes of the individual targets (just as in the case of radio ranging to the major planets).  Both astrometry and ranging require careful determination of the location of the observatory relative to the Earth-Moon barycenter.  An advantage of conducting tests in the outer solar system is that the desired gravitational signals from \px\ are meter-scale or greater, which is well above geodetic errors.  Signals at cm scales place greater stress on our knowledge of the Earth-Moon system.

\subsection{Systematic errors}

Other forms of measurement error or un-modelled sources of acceleration can potentially degrade the \px\ constraints.  If these errors are \emph{systematic}, \ie\ are correlated with the model's derivatives with respect to a free model parameter, then they 
must be controlled at the $\sigma_{\rm pop}$ level, \ie\ sub-\uas\ and sub-meter scale.  We next discuss the systematic errors that we expect to encounter, their expected sizes, and mitigation strategies.

\subsubsection{Radiation pressure}
If the ratio of incident solar radiation pressure to the solar gravitational force on a spherical target of radius $R_T$, density $\rho$, and heliocentric distance $d_T$ is well below $\epsilon,$ then we can ignore its influence.  This condition is
\begin{align}
  \epsilon & \gg \frac{\pi R_T^2 L_\odot / 4 \pi c}{4\pi GM_\odot\rho R_T^3/3} = \frac{3L_\odot}{16\pi GM_\odot \rho c}  R_T^{-1} \\
  \Leftrightarrow \quad  R_T &  \gg 17 \,\textrm{km} \times \left(\frac{5.2\,\textrm{AU}}{d_T}\right)^3
                               \left(\frac{1000\,\textrm{kg}\,\textrm{m}^{-3}}{\rho}\right).
\label{radpressure}
\end{align}
The $H<14.5$ limit of useful \lsst\ Trojans corresponds to $R_T> 4$~km at a typical albedo of 0.04 for the population \citep{Fernandez}, suggesting the smallest useful Trojans have incident radiation pressure a few times larger than the nominal tidal effect.  Intrinsic variations in albedo (and hence $R_T$) and density at fixed $H$ will make it difficult to predict this on a per-target basis. Departures from isotropy in the reflected or re-radiated energy likewise cannot be predicted accurately, \ie\ the Yarkovsky effect.  The radiation-pressure perturbation is too small, however, to be detectable for any single Trojan, and therefore we need only be concerned about the average effect over the whole population, which could indeed be statistically significant at the fainter $H$ values of the proposed sample.  When averaging over the population and over rotation phases, we would expect the dominant component to be a radial inverse-square force, which does not induce precession or quadrupole signals that would mimic \px.  It could, however, potentially confound the radial displacement signal---but we do not expect that this signal is strong enough to contribute to \px\ constraints from \lsst.  We will test this conclusion numerically by allowing the population-averaged radiation pressure to have a free nuisance parameter included in the Fisher analysis.

The angular momentum vector of a body's orbit or rotation breaks the symmetry of the system around the sun-target axis, potentially producing non-radial forces---the Yarkovsky effect.  For (101955) \textit{Bennu}, the asteroid with the best-constrained Yarkovsky models, the results of \citet{FarnocchiaBennu} are that the tangential Yarkovsky acceleration is roughly $g_Y=10\%$ of the incident radiation-pressure acceleration for this body with radius 0.25~km.  If we assume a similar $g_Y$, \ie\ anisotropy of thermal emission, then the ratio of non-radial radiation-pressure acceleration $a_Y$ to the nominal \px\ tidal acceleration becomes
\begin{equation}
  \frac{a_Y}{a_{\rm tidal}} = \frac{3 g_Y L_\odot }{16\pi R_T \rho c} \frac{d_X^3}{GM_X d^3_T} = \left(\frac{1.8\,\textrm{km}}{R_T}\right)
  \left(\frac{g_Y}{0.1}\right)
  \left(\frac{\rho}{1000\,\textrm{kg}\,\textrm{m}^2}\right)
  \left(\frac{5.2\,\textrm{AU}}{d_T}\right)^3.
  \label{eq:yarkovsky}
\end{equation}
For the smallest members of the Trojan sample, the Yarkovsky force is only a few times below the \px\ signal, and must be considered in an analysis on a population-averaged basis. Note that the population's vectorial average of $a_Y$ will be reduced because of the variation of spin axes, so the Yarkovsky signal of the Trojans could be too small to be detectable.
Two characteristics of Yarkovsky acceleration will distinguish it from \px\ perturbations.  First, if there is a tangential force, it has non-vanishing curl, \change{which is an effect that gravity cannot produce}.  More importantly, it will vary systematically with $H$ (as a proxy for $R_T$).  We will be able to regress the inferred tidal field against $H$ to determine and correct for the trend.  Introducing a nuisance parameter for the Yarkovsky effect will quantify the effectiveness of this technique.

The non-gravitational forces on free-flying spacecraft are orders of magnitude too large for use as gravity probes at our desired level.  Only spacecraft that can be referenced to the barycenters of planets (either in orbit or on fly-bys) are useful, until such time as ranging of drag-free spacecraft becomes available.

\subsubsection{Outgassing}

Another potential non-gravitational force is reflex from anisotropic mass loss (the ``rocket effect'' \citealt{whipple}).  Visible \change{comae are} very rare in both the Trojan and MBA populations.  We may safely assume that the average member of either population has existed for $>4$~Gyr and is currently losing mass at a rate below its own mass per 4 Gyr.  If we assume that the velocity of ejected mass is below 500~m/s, roughly the mean thermal velocity of water at $\approx 200$~K , then the total acceleration from outgassing or mass loss is bounded by
\begin{equation}
  \frac{g_R}{M}\frac{dp}{dt} < g_R \frac{v_{\rm thermal}} {4\,\textrm{Gyr}} \approx 4g_R \times10^{-15}\,\textrm{m}\,\textrm{s}^{-2}.
\end{equation}
Here $g_R$ is a measure of the anisotropy of the mass loss for the rocket effect, in analogy to the $g_Y$ for the Yarkovsky effect.  There will again be averaging due to the random orientation of the spin/rocket vectors among the population.
\change{A single fixed-orientation planar surface outgassing uniformly into a hemisphere would have $g_R=0.5.$ Once we vectorially average over the different orientations of patches on a given body, the rotation of the body, and the multiplicity of bodies, an average of $g_R<0.2$ seems likely, at which point} the maximal mass-loss force is $10\times$ below the nominal tidal force.
We are probably safe ignoring mass-loss forces, especially if any objects showing coma in deep \lsst\ coadded images are excluded.
    
\subsubsection{Photocenter motion}
\label{sec:pcm}
\begin{figure}
\begin{center}
  \includegraphics[width=0.5\columnwidth]{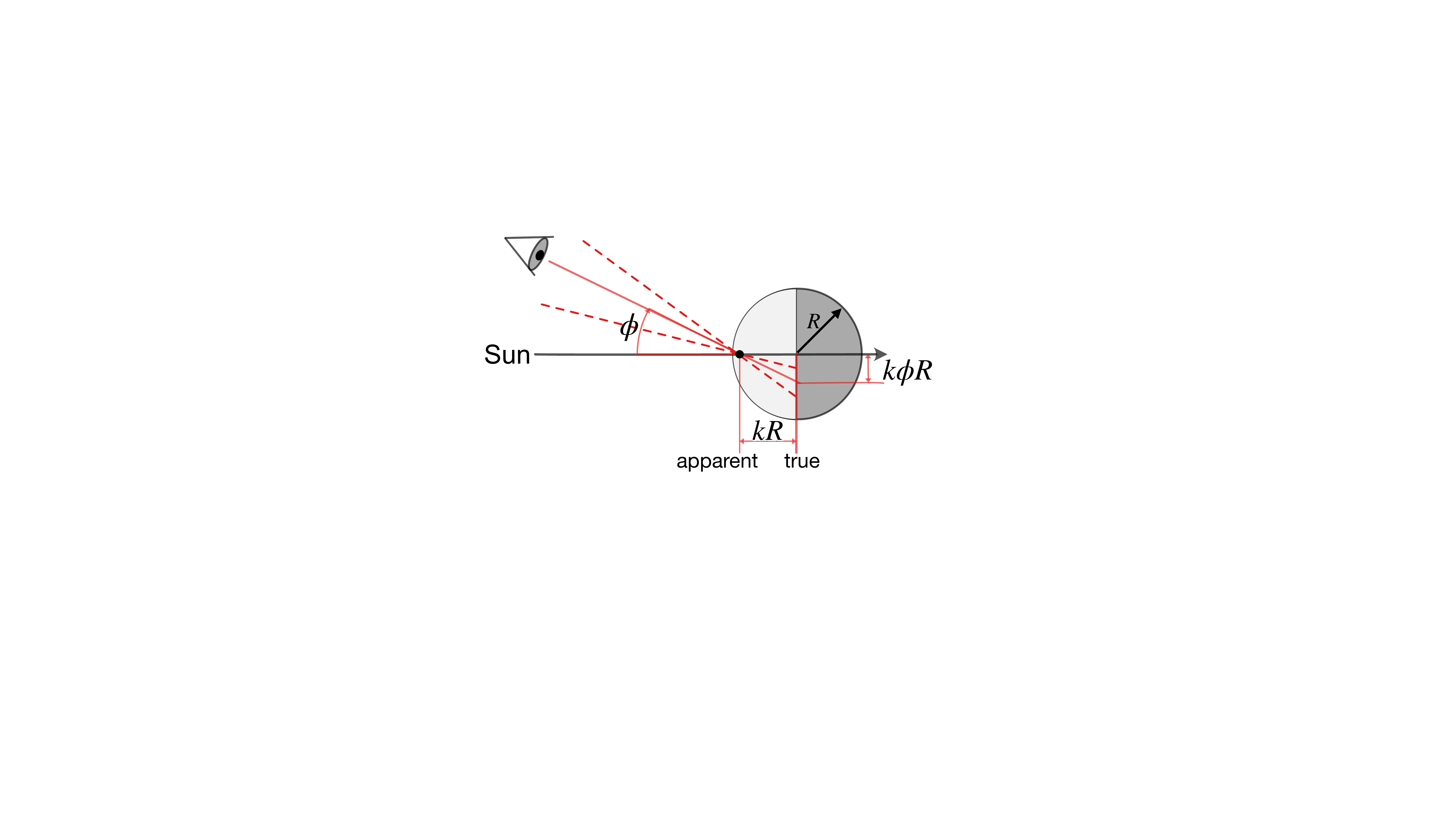}
\end{center}
\caption{If the body at right with radius $R$ is illuminated along the horizontal axis, and the photocenter displacement when viewed (in red) at an illumination angle of $\phi$ is $k\phi R,$ the naive use of the photocenter instead of the barycenter is equivalent to a shift of the body toward the Sun by $kR,$ at the location of the black dot.}
  \label{fig:pcm}
\end{figure}

The displacement between the photocenter and barycenter of a tracer becomes an error in the ephemeris calculation.  We can decompose the photocenter motion (PCM) into some apparent transverse displacement $\Delta x(\phi_0)$ that is purely a function of the Sun-Target-Observer phase angle $\phi_0$ and lies in the plane of this angle;   plus a zero-mean component that fluctuates due to the rotation of a body lacking rotation symmetry.  The latter will be uncorrelated with any gravitational signal and can be viewed as additional astrometric noise.  The former will be coherent with an oscillation at the body's synoptic period.  The phase angle $\phi_0$ is bound to $\pm\sin^{-1} \frac{1\,\textrm{AU}}{d_T}.$ In this range we will approximate
\begin{equation}
\Delta x(\phi_0) = k\phi_0 R_T
    \label{pcmeq}
\end{equation}
for some constant $k$ that will depend on the geometry and scattering properties of the tracer body.  As illustrated in Figure~\ref{fig:pcm}, the effect of this synoptic PCM is to move the apparent center of the object toward the Sun by $kR_T.$  PCM is a more severe issue for larger bodies, in contrast to radiation-pressure systematics.

The PCM shifts can be quite large.  For a specular surface, the value of $k$ is $1/2.$  Appendix~\ref{sec:lambert} shows that a Lambertian sphere has $k=3/8.$  \citet{Mallama} estimates the PCM for the Galilean satellites using photometric models of their surfaces, with results equivalent to $k\approx0.4.$  These values suggest that the angular PCM for even the smallest Trojan targets will be $\approx 4\,\textrm{km}\, \times 0.4/5.2\approx300$~m and manifest as $\gtrsim1.5$~km under-estimates of inferred radial coordinates. This is several orders of magnitude above both $\sigma_{\rm pop}$ and the nominal \px\ signatures.  Indeed our expected $\sigma_{\rm pop}$ for the Trojans is roughly $10^{-3}$ of the diameter of the smallest targets (Figure~\ref{fig:sigpop}).

We have reason, however, to be optimistic that this systematic error will not degrade the \px\ constraints.  As noted, PCM generates an apparent radial displacement, but not the precession or multipole signals that carry most of the tidal information.  Furthermore in Section~\ref{sec:sysmodels} we will examine the  impact of allowing every Trojan test particle to have a free parameter for its $k$ value.

\citet{gaiadr3ss} demonstrate direct detection of the PCM wobble at $\approx1$~mas level from the rotation of (21) \textit{Lutetia.}  PCM values this high would inflate our expected statistical error floor of $\approx2$~mas. But at $H=7.3,$ and distance 2.4~AU, \textit{Lutetia} should have a much larger angular stochastic PCM than the typical \lsst\ Trojan or MBA target.  For present purposes we can ignore the stochastic rotational portion of PCM, and concern ourselves only with the rotation-averaged values.

\subsubsection{Differential chromatic refraction}
Images of celestial objects appear closer to the zenith because of refraction in the atmosphere.  Because the refraction is wavelength-dependent and the observing bandwidth is finite, there is a differential chromatic refraction (DCR) that depends on the precise spectrum of the source.  To first order, the spectral dependence can be quantified by some broadband color measure $c_{bi}$ for each source $i$ spanning band $b.$  For a horizontally stratified atmosphere, the DCR in exposure $j$ at zenith angle $z_j$ can be quantified as an apparent shift
\begin{equation}
  \Delta z = k_{bj} c_{bi} \tan z_i
  \label{dcreqn}
\end{equation}
\change{where $k$ scales the amplitude of the effect.}
Values of $k$ measured in the $griz$ bands by \citet{decamast} are 45, 8.4, 3.2, and 1.4~mas per magnitude of $g-i$ color, respectively.  Chromaticity of the glasses in the prime-focus corrector cause an effect (``lateral color'')  of similar size and wavelength dependence,  which is oriented radially from the telescope axis.  These effects are clearly large, \ie\ $10^4\times$ larger than the target $\sigma_{\rm pop}$ for the \lsst\ Trojans.  Exquisite removal of the effect will be needed.

The DCR is easily estimated to precision $<1$~mas on every exposure by using color data for each individual target, combined with knowledge of the $k_{bj}$ gained by monitoring reference stars of varying color.  In this way we can insure that uncorrected DCR does not increase the stochastic astrometric error floor $\sigma_{\rm min}.$ The much greater challenge is in insuring that coherent residuals to the DCR corrections do not impact our inferences being made at \uas\ level from the full population.  At this level, an accurate prediction of DCR would need both precise measurement of the time-varying behavior of the atmosphere (the $k_{bi}$) and of the full spectral energy distribution of each minor planet.  It should be possible to infer the former by using the tens of thousands of \gaia\ stars that will be imaged at high $S/N$ in every \lsst\ exposure---indeed we should be able to average many exposures, since the atmospheric constituents and pressure do not change rapidly.  The $z_i$ values are of course known to very high accuracy.  The difficulty would be in knowing the relevant color factor for each body.  We will therefore include $c_{bi}$ as free parameters for every band and target when making our forecasts (Section~\ref{sec:sysmodels}).  The DCR effects can be distinguished from gravitational perturbations because the former are modulated by the pointing patterns of the telescope, \ie\ we always know that the atmospheric DCR points toward the zenith and the lateral color toward the telescope axis.  The Fisher matrix analysis will tell us how effectively this decoupling works.

\subsubsection{Timing accuracy}
The maximum apparent motion of a typical Trojan is $\approx10$~mas/s.  To keep ephemeris errors below $\sigma_{\rm pop}\approx 2\,\uas,$ any systematic errors in the temporal midpoint of the \lsst\ exposures should be $\ll 200$~microseconds.  The time to sweep the shutter across the \lsst\ focal plane is $O(1)$~s, so a very precise map of the shutter blade motion is needed.  Fortunately the \lsst\ shutter motion is tracked at $O(10)$ microsecond precision, but effort will be needed to map this onto the focal plane at this accuracy.\footnote{A. Roodman, private communication.}  The tolerance for random errors in the shutter timing is much looser; but one could imagine systematic variations that could correlate with gravitational signals. \change{For example, the gravity vector on the shutter blades could be correlated with the Trojan's position on the sky and season, since the parallactic angle between celestial north and altitude vectors at a given time of night depends on these factors.}  The faster-moving MBAs could potentially be used as calibrators of the shutter trajectory, or perhaps the rarer but even faster near-Earth asteroids.

\subsection{Full model and parameters}
\label{sec:sysmodels}
In this subsection we detail the model and parameters for the observations of the Trojans.  The techniques would be applicable to other small-body populations as well in future analyses.

For each of the Trojan asteroids, a.k.a. the targets or tracers, we begin with six free parameters: its position components $x$, $y$, and $z$, and its velocity components $v_x$, $v_y$, and $v_z$ at a reference epoch $t_0$, taken to be MJD$=60000$ (25.0 Feb 2023), near the start of the nominal LSST. We neglect the effect of individual Trojans' gravity fields on each other and on the remaining bodies. Therefore, their masses are not included as free parameters. 

\subsubsection{Trojan observations and derivatives}
The Trojans function as test particles in the modeled solar system gravitational field. The computation of the derivatives of their observable positions is done as described in Section \ref{sec:derivs}, that is, converting the state vector derivatives given by the \texttt{REBOUND} simulation into derivatives of sky coordinates.

\subsubsection{Photocenter motion}
The photocenter motion displacement is computed from  \eqq{pcmeq}, with its direction given by the phase position angle $\phi_{PA}$. This gives us
\begin{equation}
\Delta \delta = k  R_T \phi_0 \cos{\phi_{PA}}; \qquad \Delta \change{(\alpha \cos{\delta})} = k  R_T \phi_0 \sin{\phi_{PA}}
    \label{pcmmodel}
\end{equation}
The $kR_T$ term is our free parameter, and derivatives with respect to it are included in our Fisher matrix modeling for each Trojan.
\subsubsection{Differential chromatic refraction}
We model the differential chromatic refraction according to \eqq{dcreqn}, with values of $k_b$ taken from \citet{decamast}, since we assume correction over atmospheric conditions can remove the systematic effect of different observational conditions. The direction of the displacement is given by the parallactic angle $p$, such that its components are
\begin{equation}
\Delta \delta = k_b c_{bi} \tan{z_i}\cos{p}; \qquad \Delta \change{(\alpha \cos{\delta})} = k_b c_{bi} \tan{z_i} \sin{p}
    \label{dcrmodel}
\end{equation}

The zenith and parallactic angles are computed from the simulated observations. We end up with four color parameters $c_{bi}$ \change{(one per filter)} for each Trojan $i$, each of them with derivatives to be included in the Fisher matrix. 

\subsubsection{Radiation pressure}
The radial acceleration $a_R$ due to solar radiation pressure is considered to be proportional to $A_r/a^2$, with $A_r$ defined in Equation \ref{eqrad} and $a$ being the semi-major axis. 

\begin{equation}
\label{eqrad}
A_r = \frac{L_\odot R_T^2}{4 Mc}
\end{equation}

The exact proportionality depends on the albedo and shape of the object. Assuming a constant density for all Trojans, the proportionality becomes $a_R \propto c_{\rm RP}/(a^2R_T)$, where $c_{\rm RP}$ is now a nuisance parameter that scales the whole population's radiation pressure force. The derivatives of the asteroid state vectors with respect to $c_{\rm RP}$ are proportional to derivatives with respect to $M_\odot$. We obtain the former by rescaling the latter according to the radius $R_T$ of each Trojan, which is inferred from its magnitude $H$.

\subsubsection{Yarkovsky effect}

Finally, we consider the influence of the Yarkovsky effect on the asteroids.  As summarized in \citet{Bottke_2006}, the Yarkovsky effect can be separated into two components, known as the seasonal effect (which acts along the spin axis) and diurnal effect (which acts in a plane orthogonal to the spin axis).  For a population of asteroids with randomly oriented spin axes, there will be no net effect from the diurnal effect.  For the same population the seasonal effect produces a net acceleration tangent to the orbit.  The exact acceleration experienced by an asteroid is a complicated function of its shape, albedo, and orbit, but since our derivatives are insensitive to exact value assumed, an order of magnitude estimate will suffice.  If the bodies are large (with diameters larger than $\approx 50$ meters) the thermal acceleration can be computed using Equation~(16) in  \citet{Vokrouhlicky_1998}
\begin{equation}
    a_T = - \frac{4 \alpha}{3} \frac{A_Y}{a^2} A(\ell),
    \label{eq:yarkovsky2}
\end{equation}
where $\alpha$ is the albedo, $a$ is the semi major axis in AU, and $A(\ell)$ a function of the temperature distribution.  $A_Y$ sets the scale of the effect, and is a function of the mass and radius of the object being irradiated. The overall scaling $A_Y$ is taken as a nuisance parameter.  The derivative with respect to $A_Y$ of each Trojan's position is determined numerically  by adding to each test particle in the simulation a small $\hat{\theta}$ acceleration with magnitude of $10^{-12} \mathrm{m}\,\mathrm{s}^{-2}$, a rough value for a body with a radius of $50$~m and a density of $\rho=3500\mathrm{kg}\,\mathrm{m}^{-3}$. Derivatives of each body's state vector with respect to this nominal acceleration are rescaled to derivatives with respect to $A_Y$ using \eqq{eq:yarkovsky2}.

\subsubsection{Parameter summary}
There are 11 parameters for each Trojan---six phase space components, four coefficients of DCR $c_{b}$, and a coefficient $k$ for the photocenter motion.  At a global level, there remain \change{the masses of the gravitationally active bodies, along with the solar $J_2$ moment and the state vector parameters of the major planets,} plus scaling parameters $c_{RP}$ and $A_Y$ for the radial and Yarkovsky components of radiation pressure.

The total dimensionality of the Fisher matrix is thus $O(10^5)$ once all the Trojans are included.  The matrix is, however, highly structured, which allows a huge acceleration of the marginalization over nuisance parameters.  We can divide the parameter set into a group $p_k$ of 11 for the $k$th test body (Trojan), and then a group $p_G$ of ``global'' parameters comprising the specifications of all the active bodies, plus the global systematic parameters for the Trojans.  When ${\mathcal F}$ is broken into blocks among these parameter subgroups, all of the blocks representing covariances between $p_k$ and $p_{k^\prime}$ are zero for $k\ne k^\prime$. This means that  \eqq{marginalization2} can be successively applied to all Trojans to marginalize down to the global parameters:
\begin{equation}
  {\mathcal F} \rightarrow {\mathcal F}_{GG} - \sum_k {\mathcal F}_{Gk} {\mathcal F}_{kk}^{-1} {\mathcal F}_{kG}.
\end{equation}
The marginalization over all Trojans' parameters never requires inversion of matrices larger than $11\times11$ and there is never need to store the full Fisher matrix.

\section{Results with Trojan tracking}
\label{sec:trojanresults}
\subsection{Combinations with ranging data}
\label{sec:combinations}
We now calculate the full-sky Fisher uncertainties $\sigma_M$ on \px\ mass in scenarios 
including predicted \lsst\ observations of Jovian Trojans.  The dashed lines in Figure~\ref{skyfrac} plot the sky fractions achieving a given $\sigma_M$ level when the \lsst\ data are combined with the nominal major-planet data, in the four possible choices of whether the Juno/Cassini and Mars data are included.  The summary statistics for these four scenarios are in the lower half of Table~\ref{limitstab}.

The addition of the \lsst\ astrometry for Trojans to the full Old$+$J/C$+$Mars nominal observation set yields very minor improvements in \px\ detectability.  The biggest gain occurs near the ecliptic poles, where addition of the Trojans lowers $\sigma_M$ from 1.07 to $0.98M_\oplus.$  The Trojan information is, however, more useful if we have dropped one or both of the J/C and Mars ranging sets from the constraints. Thus the \lsst\ Trojans could be valuable in building confidence in a detection from the nominal data by confirming that the detection is robust to deletion of Mars or J/C data.

Adding the Trojans to the Old data reduces the median $\sigma_M$ on the sky from 12.4 to 2.2$M_\oplus.$  Thus a $10M_\oplus$ \px\ would become securely detectable over 20\% of the sky completely independent of the modern ranging data.
  
Adding \lsst\ Trojan data to the Old$+$Juno/Cassini data raises the portion of the sky amenable to $5\sigma$ detection of $M_X=5M_\oplus$ at 400~AU from 87\% to 89\%.  The lower panel of Figure~\ref{maps} shows this improvement relative to the middle panel.  One sees that the Trojans are particularly valuable in filling in the weak coverage near the ecliptic poles.  When Trojan data are added to Old$+$Mars data, they improve the sensitivity to a polar \px\ by a factor of 2.
  
\subsection{Effect of $\sigma_{\rm min}$}
\label{sec:varySigmaMin}  

A reduction on the turbulence floor $\sigma_{\rm min}$ of the \lsst\ observations from 2~mas to 0.5~mas does not yield drastic improvement on constraining power in the nominal case with Trojans.  The $\sigma_M$ near the ecliptic poles is decreased from 0.98 to $0.89M_\oplus$, with lesser gains elsewhere on the sky.  
This suggests that an astrometric limit of 2~mas already extracts most of the \px\ information from the \lsst\ exposures. 

\subsection{Suppression of systematic errors}
\begin{figure}
 \centering
 \includegraphics[width=0.9\textwidth]{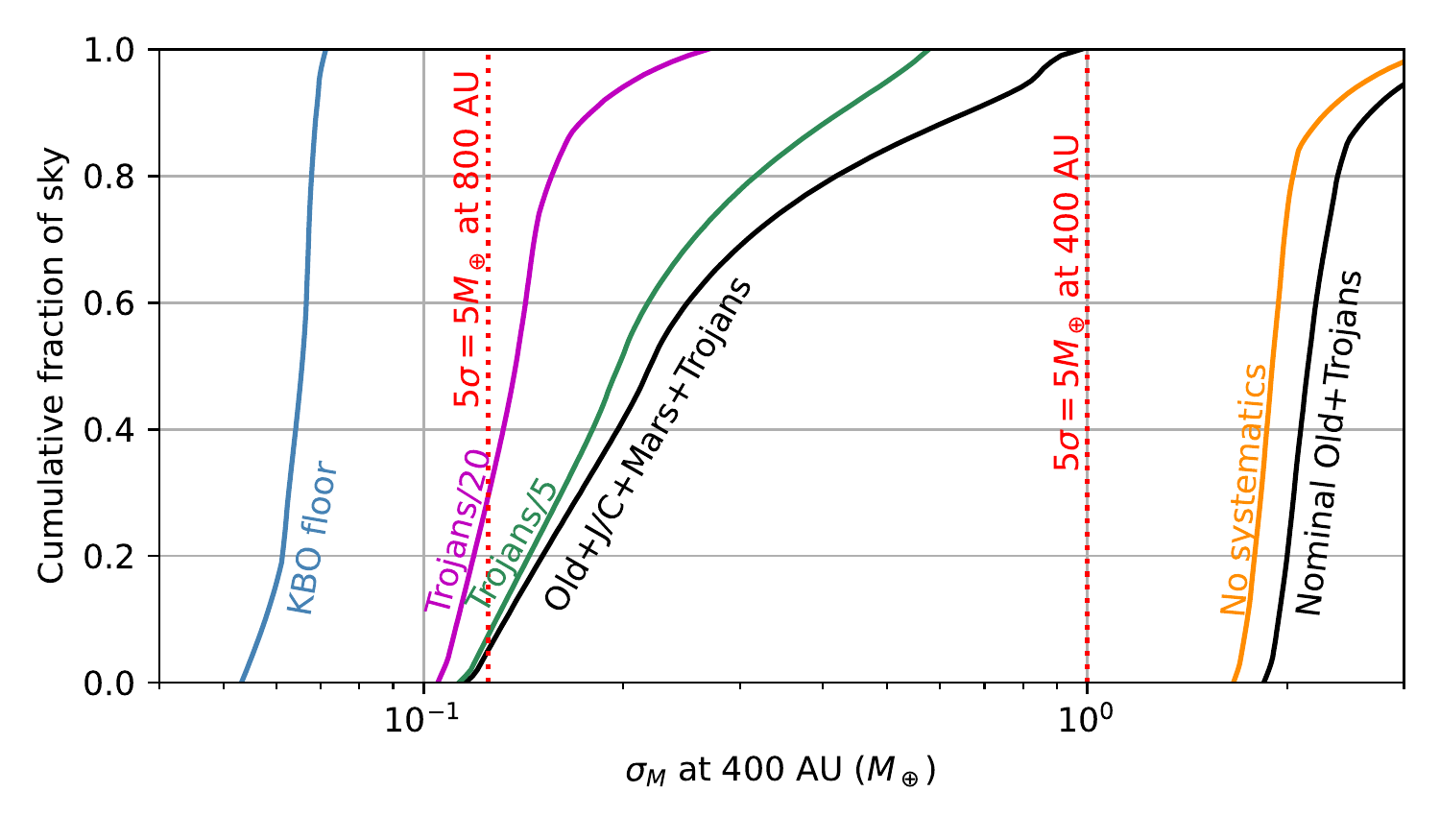}
 \caption{The fraction of the sky with uncertainty $\sigma_M$ on \px\ mass below a given level is shown for observation scenarios including \lsst\ observations of Trojans.  At the right end we see the relatively small degradation in the combination of Trojans with old data that results from marginalizing over parameters of systematic-error models \change{(as we move from the yellow to the dark curve on its right)}.  At left are shown scenarios in which we combine all of the proposed observational data, and arbitrarily reduce all uncertainties on the Trojans by factors of 5 (green) and 20 (magenta) from the actual forecast (black).  It takes the 20-fold reduction for the Trojans to make a substantial improvement in the \px\ constraints, and gain the ability to detect a $5M_\oplus$ \px\ at $5\sigma$ significance over a significant portion of the sky if it is \change{at a distance of} 800~AU. At far left, below the 800~AU signal, is the floor set by stochastic KBO quadrupoles.}
 \label{skyfrac_variants}
\end{figure}

The observational and physical Trojan systematic error parameters can be crossed out of the Fisher matrix in order to generate an ideal case where these parameters are fully known, \ie\ there are no systematic errors. With this test, we can examine the level to which the systematic effects are degenerate with a \px\ signal.  For this test we compare the Old$+$Trojans scenario with and without the systematic error parameters free to vary, omitting the J/C$+$Mars ranging to maximize the potential harm from Trojan systematics.  The rightmost two curves in Figure~\ref{skyfrac_variants} plot this comparison, where we find 
a reduction of 12\% in the median value of $\sigma_M$ when these effects are suppressed. 
Even though some of the systematic signals are much larger than the \px\ perturbations, we can successfully exploit their known dependence on target size or observing geometry to \change{yield} only  minor degradations in \px\ precision relative to the no-systematics case.

\subsection{Avenues for stronger constraints}

A fundamental limit in searching for the tidal field from unknown solar system masses is the degeneracy with the tidal field produced by \emph{known} elements of the solar system, namely Kuiper Belt members whose masses are poorly known, and whose periods are so long that we cannot use time dependence on decade scales to distinguish them from \px. In our Fisher matrix we include these uncertainties in three forms.  First, there is an uncertainty in the azimuthally symmetric ``ring'' of the time-averaged Kuiper belt.  This we find to have relatively little effect on $\sigma_M,$ perhaps because it only generates the $m=0$ spherical harmonic, and any non-polar \px\ will have components at $m=\pm1$ and $m=\pm2$.  Second, we have included the tidal fields from the largest individual KBOs.  These are substantially reduced by the fact that most large TNOs have moons, and hence we can obtain uncertainties on their masses at $O(10^{-6}M_\oplus).$  Since the tidal field scales as $M_X/d_X^3,$ and the KBOs are at $d\ge40$~AU while we are calibrating $\sigma_M$ for $d_X=400$~AU, we expect the $\approx 10^{-5}M_\oplus$ uncertainties in the masses of the largest moon-less KBOs to result in an $\approx 10^{-2}M_\oplus$ floor on $\sigma_M.$
We also note that the tidal fields scale as $d^{-3}$ from their source, so additional large TNOs that may exist beyond the Kuiper Belt ($d\gtrsim50$~AU) are of relatively little importance.  
We find the dominant contribution to the KBO floor, 
plotted in the leftmost curve in Figure~\ref{skyfrac}, is the third component, from stochastic fluctuations in the quadrupole moment of the large numbers of smaller KBOs that occur as individual elements orbit the Sun.  Reducing this uncertainty will unfortunately prove more difficult than quantifying only the few largest KBOs, since it would require discovering and making reasonable mass estimates for many objects all over the sky, something even \lsst\ will not do.

The good news is that this KBO floor will not be a problem until the data get substantially better. The KBO floor at $\sigma_m\approx0.06M_\oplus$ is about 3 times smaller than the uncertainties forecast to be available from observations in the coming decade even in the most favorable directions, so there is room for improvement with better data.  Furthermore the KBO floor lies below the accuracy required to find the tidal field of a $5M_\oplus$ \px\ at $d_X=800$~AU anywhere on the sky.

What observations would be needed to bring the achievable
$\sigma_M$ constraints significantly closer to the KBO floor?
The three middle curves of Figure~\ref{skyfrac_variants}  (also in Table~\ref{limitstab}) consider cases in which we arbitrarily reduce all of the measurement errors on \lsst\ Trojans.  It is apparent that the overall sensitivity to \px\ begins to significantly improve when these errors are reduced to $\lesssim0.1$ those achievable with \lsst.  At $20\times$ reduced errors, we can confidently detect the most pessimistic version of \px, $5M_\odot$ at 800~AU, suggested by dynamical studies, over 29\% of the sky. The measurement uncertainties would still be significantly above the KBO floor.

These 10--20$\times$ reductions over \lsst\ capabilities are potentially achievable over a few decades' observations with an \lsst-like facility, if its observing time is more strongly concentrated on the Trojans instead of the full sky, and if we note that the size of the tidally-induced precessions grow linearly with time span.  Ranging to future spacecraft missions to the giant planets would obviously be of substantial gain, as would the advent of ranging to drag-free spacecraft, \eg\ with cm-scale accuracy to a test body beyond Jupiter.

A less expensive route to higher accuracy on minor-planet positions is to implement the stellar occultation detection array proposed by \citet{RiceLaughlin}, which can potentially reduce per-epoch positional uncertainties from milliarcseconds to $O(10)$~\uas\ for occultations of brighter \gaia\ stars.  We leave a detailed quantitative study of this possibility for future work.

\section{Conclusions}
\label{sec:conclusions}

Using Fisher matrix analysis, we find that current tracking of major planets through range and astrometric measurements has achieved a precision such that the tidal effect of an unknown Planet X can be distinguished from spurious signals throughout most of the sky. Ranging data from Juno, Cassini and Mars orbiters contribute most to the total Fisher information. In full-sky searches, the fraction of the sky where the Cramér-Rao bound allows for a $5\sigma$ detection of a $5M_\oplus$ \px\ is $f_{400}=99.2\%$ for \px\ at 400 AU and $f_{800} = 4.8\%$ at $800 AU$.

How do actual attempts to detect \px\ in the data compare to our derived bounds?
Previous gravitational searches are still less conclusive than this prediction. The strongest published bounds are from
\citet{fienga2020}, who perform a gravitational search for \px\ by perturbing  the INPOP19a ephemeris. Their most readily interpretable results are likelihoods of the $\Delta\chi^2$ of ephemeris best fits with an added \px\ tidal field to those from a null hypothesis with no \px\ present.  \citet{fienga2020} test cases of $M_X=5$ and $10M_\oplus$ placed at $400<d_X<800$~AU, at ecliptic latitudes between $\pm25\arcdeg.$  Comparison with our forecasts is a little clouded because they do not search the full sky, and their results do not display the inversion symmetry and the $M_X/d_X^3$ scaling expected for a tidal force [our \eqq{eq:tidal} and their Equation~(1)], implying that some numerical noise is present.
Their Figure~5 and text suggest that a $5M_\oplus$ \px\ at 600~AU is near a $3\sigma$ detection in most of their footprint, equivalent to $\sigma_M\approx 0.5M_\odot$ in our reference case of $d_X=400~AU$.  A similar conclusion is drawn from their depiction of $\approx1\sigma$ signals from $M_X=5M_\oplus$ at 800~AU.  This constraint is 2 times weaker than the median $\sigma_M\ge0.24M_\oplus$ bound that we derive for the comparable case of using published ranging in Table~\ref{limitstab}.  Thus their results do not violate our bound, and it appears that stronger \px\ constraints should be possible with the available data.
They remark that \px\ perturbations were often compensated in the fitting process by changes to the TNO ring mass, which points to a degeneracy between these parameters. Our analysis shows, however, that a \px\ signal can be sufficiently distinguished from a symmetric KBO ring with current major planet data---suggesting that this is not currently a limiting degeneracy.  We find that quadrupolar asymmetries within the ring are the stronger degeneracy with \px, but such perturbations are not present in the \citet{fienga2020} model. \change{Also of note is the bound from \citet{Meisner_2017} which ruled out a Planet X brighter than $W1 = 16.66$ from a search of the \textit{WISE} spacecraft images.  Using the models of \citet{Fienga_2016}, this $W1$ limit corresponds roughly to excluding $M_X\geq 20 M_\oplus$ at $622$~AU. Our forecasted gravitational bounds on $M_X$ are tighter than this direct-imaging result.} From this somewhat confusing state of affairs, we conclude that further dynamical fitting and modelling is of substantial value.

We further investigate whether the predicted observations of Jovian Trojans with the upcoming LSST survey would improve these constraints. The answer is that, when all data on the major planets are considered, not much information would be gained from these additional measurements, and a 10--20$\times$ improvement in their predicted uncertainty would be needed before they are useful in this scenario. However, when different subsets of the major planet data are ignored, there is a considerable gain of information from LSST Trojan tracking. Therefore, these measurements would be useful as cross-checks for the results obtained with specific major planet data sets. Our calculations also indicate that \change{astrometric data for} a large population of \change{small} bodies can internally solve for (and remove) the major sources of systematic error, attaining the full meter-scale total statistical accuracy for true gravitational signals.

Our discussion has been framed around the search for a single \change{sub-Neptune-mass} planet lurking $\ge400$~AU away, but the tidal-field constraints are equally applicable to smaller or multiple bodies that are $\gtrsim60$~AU away such that their gravitational influence is dominated by a tidal field.  For our nominal case of using extant data to constrain the tidal field's components, Figure~\ref{fig:fosky} indicates that $\sigma_M<0.7M_\odot$ for a single mass at 400~AU over 90\% of the sky.  Using the scaling of \eqq{eq:tidal}, this indicates that an unknown Earth-mass planet should be detectable at $5\sigma$ significance for $d_X\le 260$~AU, and a Mars-mass object for $d_X\le120$~AU.  There is currently no evidence to exclude the existence of such bodies. The scaling of a gravitational search for new masses, with signal $\propto d_X^{-3},$ is more favorable than for searches for reflected solar light, which is $\propto d_X^{-4}.$ 

The $5\times5$ Fisher matrix we have derived for the tidal field is equally applicable to cases in which a tidal field is generated from \emph{multiple} undiscovered bodies. Investigating the bounds on cases of multiple distant planets would be a straightforward extension.  Note that only a subset of the space of 5-dimensional tidal fields is compatible with generation from a single point mass, so there is a potential internal test for the multiplicity of sources.  Likewise one can estimate from the Fisher matrix our future capability to localize any single source on the sky given its mass and distance.

\change{Performing these localizations will be computationally intensive in practice.  However, since the computation of the variational derivatives can be parallelized, we anticipate no major computational bottlenecks associated with processing a large future dataset. The principle challenge will be incorporating physics with sufficient accuracy to compare simulations with observations. Necessary improvements would include the $J_2$ geopotential terms for several planets and a precise treatment of general relativistic perturbations. As far as we know, no open source pipeline currently exists that could preform this integration, so one will have to be developed for this purpose. 
}

Our results indicate which steps can be taken towards a better detection of the \px\ signal. Fits of the major-planet data to integrated ephemerides should be capable of stronger constraints than have been obtained, motivating investigation of integrator accuracy and fitting techniques. Improved efficiency for such integration and fitting processes could be needed.  Further characterization of the potential roles of other minor planets (\eg\ MBAs or Hildas) as test bodies is warranted. In the longer term, there is significant improvement possible in our sensitivity to \px\ from spacecraft visiting the giant-planet systems, ranging to drag-free spacecraft, and campaigns to measure many thousands or millions of stellar occultations by minor planets.

\begin{acknowledgments}
The authors thank the NSF Astronomy \& Astrophysics Program for supporting this work: grant AST-2206194 for MJH and ZM, and grants AST-2205808 and AST-2009210 for GMB, DCH, and RCH.  MJH is also grateful for support from the Smithsonian Scholarly Studies Program and from the NASA YORPD Program (80NSSC22K0239). We greatly thank the members of the \lsst\ Survey Strategy Optimization Committee, in particular Lynne Jones, for assembling the simulated \lsst\ surveys and helping us to use them properly.
\end{acknowledgments}

\bibliography{refs}{}

\appendix
\section{Tidal field decompositions}
\label{sec:tidal}
Let $\phi$ be the gravitational field at position $r$ generated by a body at $r'$,
\begin{equation}
    \phi=-\frac{GM}{|r-r'|}.
\end{equation}
For $r<r'$, that is, the inner field,
\begin{equation}
    \frac{1}{|r-r'|}= \sum_{\ell=0}^\infty \frac{4\pi}{2\ell+1}\sum_{m=-\ell}^\ell (-1)^m\frac{r^\ell}{r'^{\ell+1}}Y_\ell^{-m}(\theta,\phi)Y_\ell^m(\theta',\phi');
\end{equation}
therefore, the quadrupole term is given by
\begin{equation}
    Q(M,\vec{r},\vec{r'}) = -\frac{4\pi GMr^2}{5r'^3}\sum_{m=-2}^{2}(-1)^mY^{-m}_2(\theta,\phi)Y_{\change{2}}^m(\theta',\phi').
\end{equation}
Let $K(r,r') = -4\pi Gr^2/(5r'^3)$,
\begin{equation}
\label{quad_expansion}
    Q(M,\vec{r},\vec{r'}) = K(r,r')\sum_{m=-2}^{2}Mc_m(\theta',\phi') Y^{m}_2(\theta,\phi)
\end{equation}
where $c_m(\theta',\phi') = (-1)^mY_2^{-m}(\theta',\phi')$.

Let us define five point masses at the same distance from the origin $|r'|$, whose quadrupole fields are linearly independent:
\begin{equation}
    Q_i(M_i,\vec{r},\vec{r_i}) = K(r,r')\sum_{m=-2}^2 M_i c_{m,i}(\theta_i,\phi_i)Y^{m}_2(\theta,\phi).
\end{equation}

We can express the quadrupole field of a point mass at any other angular position as a linear combination of these five nominal fields:

\begin{equation}
    Q(M,\vec{r},\vec{r'}) = \sum_{i=0}^5 a_i(\theta',\phi')Q_i(M_i,\vec{r},\vec{r_i})
\end{equation}
\begin{equation}
   \Rightarrow \  Q(M,\vec{r},\vec{r'}) = K(r,r')\sum_{m=-2}^2\sum_{i=0}^5 M_i a_i(\theta',\phi')c_{m,i}(\theta_i,\phi_i)Y^{m}_2(\theta,\phi). 
\end{equation}
From \eqref{quad_expansion},
\begin{equation}
    Mc_m(\theta',\phi') = \sum_{i=0}^5 M_ia_i(\theta',\phi')c_{m,i}(\theta_i,\phi_i).
\end{equation}
If we assume the mass is equal for all possible angular positions,
\begin{equation}
\label{eq_for_vector_a}
    c_m(\theta',\phi') = \sum_{i=0}^5 a_i(\theta',\phi')c_{m,i}(\theta_i,\phi_i).
  \end{equation}

The vector $a(\theta',\phi')$ for a given direction in the sky is found by solving Equation \eqref{eq_for_vector_a}, and can be used to contract the 5-dimensional Fisher matrix,
\begin{equation}
    a^T \mathcal{F}a = 1/\sigma_M^2.
\end{equation}

This allows us to find the Fisher information on the mass of \px\ at any position using derivatives computed solely for five positions.
\section{Optimal astrometric accuracy}
\label{sec:astrometry}
When fitting a model of a point-spread function (PSF) to an unresolved image of an asteroid, there are three free parameters: flux $f$, and the centroids $x_0,y_0.$  The expected signal $f_k$ in pixel $k$ at location $(x_k,y_k)$  subtending solid angle $\Delta x \Delta y$ is
\begin{equation}
   \bar f_k   = \left[b + f P(x_k-x_0,y_k-y_0)\right] \, \Delta x \, \Delta y 
 \end{equation}
 where $P$ is the PSF function, we assume that the sky background $b$ is in units of photocarriers per unit solid angle, the flux $f$ is in units of photocarriers.  If the measurement $f_k$ is distributed as $f_k \sim \textrm{Poisson}(\hat f_k),$ then it is straightforward to show that the Fisher matrix in \eqq{eqn:fisher} for $f_k$ is
 \begin{equation}
   {\mathcal F}_{ij}^k = \frac{1}{\bar f_k} \frac{\partial \hat f_k}{\partial p_i}  \frac{\partial \hat f_k}{\partial p_j}
 \end{equation}
 and the full Fisher matrix is the sum over the statistically independent pixels $k$.  For a circularly symmetric PSF $P(r)$, we can combine the above 2 equations and use the parameter vector $\{f,x_0,y_0\}$ to obtain
 \begin{align}
   {\mathcal F}_{ff} & = \sum_k \Delta x\, \Delta y\,\frac{  P^2(r_k) } {b + fP(r_k)} \rightarrow 2\pi \int r\,dr\, \frac{  P^2(r) } {b + fP(r)} \\
   {\mathcal F}_{xx} =  {\mathcal F}_{yy}  & = \sum_k \Delta x\, \Delta y\,
                                             \frac{  f^2 } {b + fP(r_k)} \left[ \frac{\partial P(r)}{\partial r}\right]^2_k\cos^2 \theta_k
                                             \rightarrow \pi f^2 \int r\,dr\, \frac{ 1 } {b + fP(r)} \left[ \frac{\partial P(r)}{\partial r}\right]^2
\label{eqn:fxx}
 \end{align}
 with the off-diagonal elements being zero.  The uncertainty  $\sigma_x$ in the $x$ position is equal to that in the $y$ direction and is the inverse square root of ${\mathcal F}_{xx}.$  If we take the PSF to be a circular Gaussian
 \begin{equation}
   P(r) = \frac{e^{-r^2/2\sigma_{\rm PSF}^2}}{2\pi\sigma^2_{\rm PSF}},
 \end{equation}
 then we can change \eqq{eqn:fxx} to be a dimensionless integral
 \begin{align}
   \label{eq:sigmax}
   \sigma^{-2}_x & = \frac{f}{\sigma^2_{\rm PSF}} \int_0^{\rm \infty} u\, du\, \frac{e^{-2u}}{e^{-u} + \nu}, \\
   u & \equiv \frac{r^2}{2\sigma^2_{\rm PSF}} \\
   \nu & \equiv \frac{2\pi \sigma^2_{\rm PSF} b}{f}.
 \end{align}
 The parameter $\nu$ is the ratio of background flux to source flux.
 
\section{Photocenter Displacement}

\label{sec:lambert}
We derive here an expression for the photocenter of a body.  Let the body be centered at the origin of a spherical coordinate system $(r,\theta,\phi)$ with the observer direction \ohat\ along $\theta=\pi/2, \phi=0.$  Let $A$ denote a position on the visible surface of the body, with $\nhat(A)$ the normal to that surface.  We will assume that the intensity of light emitted from the surface is azimuthally symmetric about \nhat, so we can denote it as $I[A, \nhat(A)\cdot \ohat].$  The photocenter shift in the direction $\yhat = (\theta=\pi/2, \phi=\pi/2)$
\begin{align}
  \bar y & =   \frac{ \int dA\,  \left[\nhat(A)\cdot \ohat\right] \times I\left[A,  \nhat(A)\cdot \ohat\right]  \times y(A)}
           {\int dA\,  \left[\nhat(A)\cdot \ohat\right] \times I \left[A,  \nhat(A)\cdot \ohat\right] } \\
         & = \frac{ R^3 \int_{0}^{\pi} \sin \theta\, d\theta \int_{-\pi/2}^{\pi/2}d\phi\, \sin\theta \cos\phi \, I\left[(\theta,\phi), \sin\theta \cos\phi\right] \sin\theta\sin\phi}
           { R^2 \int_{0}^{\pi} \sin \theta\, d\theta \int_{-\pi/2}^{\pi/2}d\phi\, \sin\theta \cos\phi \, I\left[(\theta,\phi), \sin\theta \cos\phi\right]} \\
\end{align}

where the second line assumes a spherical surface.  Next we can specify a Lambertian reflecting surface, illuminated from the (solar) direction $\shat=(\theta=\pi/2, \phi=\phi_s)$ with a flux $f_s,$ for which the scattered intensity is isotropic $I \left[A,  \nhat(A)\cdot \ohat\right]=(\nhat\cdot\shat) f_s/ \pi = f_s\sin\theta\cos(\phi-\phi_s)$.  Remembering to include only the illuminated hemisphere in the $\phi$ integration, we obtain
\begin{align}
  \bar y & =  \frac{ f_s R^3/\pi  \int_{0}^{\pi} \sin^4 \theta\, d\theta \int_{-\pi/2+\phi_s}^{\pi/2}d\phi\, \cos \phi \cos(\phi-\phi_s) \sin\phi }
           { f_s R^2/\pi  \int_{0}^{\pi} \sin^3 \theta\, d\theta \int_{-\pi/2+\phi_s}^{\pi/2}d\phi\, \cos\phi \cos(\phi-\phi_s) } \\
         &  = \frac{ (f_sR^3/16) ( 2 \sin\phi_s +  \sin 2\phi_s)}
           { (2f_sR^2 / 3\pi)\left[(\pi-\phi_s)\cos\phi_s + \sin\phi_s\right]} \\
         & = \frac{3\pi R}{16} \sin\phi_s \left[ \frac{1 + \cos\phi_s} {(\pi-\phi_s)\cos\phi_s + \sin\phi_s}\right] \\
    & \approx \frac{3}{8} \phi_s R \qquad (\phi_s < 1). \\
\end{align}
The denominator of the second line is the phase function for the Lambertian sphere.  The final result agrees with that given by \citet{AksnesPCM} in a more complicated derivation.
\end{document}